# Revealing the Mysteries of Venus: The DAVINCI Mission

James B. Garvin[1], Stephanie A. Getty[1], Giada N. Arney[1], Natasha M. Johnson[1], Erika Kohler[1], Kenneth O. Schwer[1], Michael Sekerak[1], Arlin Bartels[1], Richard S. Saylor[1], Vincent E. Elliott[1], Colby S. Goodloe[1], Matthew B. Garrison[1], Valeria Cottini[2], Noam Izenberg[3], Ralph Lorenz[3], Charles A. Malespin[1], Michael Ravine[4], Christopher R. Webster[5], David H. Atkinson[5], Shahid Aslam[1], Sushil Atreya[6], Brent J. Bos[1], William B. Brinckerhoff[1], Bruce Campbell[7], David Crisp[5], Justin R. Filiberto[8], Francois Forget[9], Martha Gilmore[10], Nicolas Gorius[1], David Grinspoon[11], Amy E. Hofmann[5], Stephen R. Kane[12], Walter Kiefer[13], Sebastien Lebonnois[9], Paul R. Mahaffy[1], Alexander Pavlov[1], Melissa Trainer[1], Kevin J. Zahnle[14], Mikhail Zolotov[15]

[1] NASA Goddard Space Flight Center, Greenbelt, MD 20771 USA

[2]Agenzia Spaziale Italiana, Rome, Italy

[3]Applied Physics Lab, Johns Hopkins University, Laurel, MD 20723 USA

[4]Malin Space Science Systems, San Diego, CA 92191 USA

[5] Jet Propulsion Laboratory, California Institute of Technology, Pasadena, CA 91109

[6]University of Michigan, Ann Arbor, MI 48109 USA

[7]Smithsonian Institution, Washington, D.C., 20560 USA

[8]NASA Johnson Space Center, Houston, TX 77058 USA

[9] Laboratoire de Météorologie Dynamique/IPSL, Sorbonne Université, ENS, PSL Research University, Ecole Polytechnique, CNRS, Paris France

[10]Wesleyan University, Middletown, CT 06459 USA

[11]Planetary Science Institute, Tucson, AZ 85719 USA

[12]University of California Riverside, Riverside, CA 92521 USA

[13]Lunar and Planetary Institute/USRA, Houston, TX 77058 USA

[14]NASA Ames Research Center, Moffett Field, CA 94035 USA

[15]Arizona State University, Tempe, AZ, 85287 USA

Abstract

The Deep Atmosphere Venus Investigation of Noble gases, Chemistry, and Imaging (DAVINCI) mission described herein has been selected for flight to Venus as part of the NASA Discovery Program. DAVINCI will be the first mission to Venus to incorporate science-driven flybys and an instrumented descent sphere into a unified architecture. The anticipated scientific outcome will be a new understanding of the atmosphere, surface, and evolutionary path of Venus as a possibly once-habitable planet and analog to hot terrestrial exoplanets. The primary mission design for DAVINCI as selected features a preferred launch in summer/fall 2029, two flybys in 2030, and descent sphere atmospheric entry by the end of 2031. The *in situ* atmospheric descent phase subsequently delivers definitive chemical and isotopic composition of the Venus atmosphere during a cloud-top to surface transect above Alpha Regio. These *in situ* investigations of the atmosphere and near infrared descent imaging of the surface will complement remote flyby observations of the dynamic atmosphere, cloud deck, and surface near infrared emissivity. The overall mission yield will be at least 60 Gbits (compressed) new data about the atmosphere and near surface, as well as first unique characterization of the deep atmosphere environment and chemistry, including trace gases, key stable isotopes, oxygen fugacity, constraints on local rock compositions, and topography of a tessera.

## 1. Introduction

The atmosphere of Venus atmosphere holds clues to its origin, evolution, and dynamics and may reflect the history of putative past oceans and active volcanism (Bougher et al. 1997; Crisp et al. 2002; Treiman 2007; Baines et al. 2013; Glaze et al. 2017; 2018; Garvin et al. 2020a; 2020b; D'Inecco et al. 2021). The selected DAVINCI mission (Figure 1) described herein responds to major lingering questions about Venus, consistently prioritized by Venus Exploration Analysis Group (VEXAG) documents (O'Rourke et al. 2019) and the 2012 Planetary Decadal Survey (NRC 2011). The mission consists of a carrier relay imaging spacecraft and a descent sphere that will be dropped into the atmosphere above Alpha Regio, an enigmatic *tessera* (i.e. mountainous, strongly tectonically deformed highland) terrain whose composition may reflect remnants of ancient continental crust (Hashimoto et al. 2008; Gilmore et al. 2015).

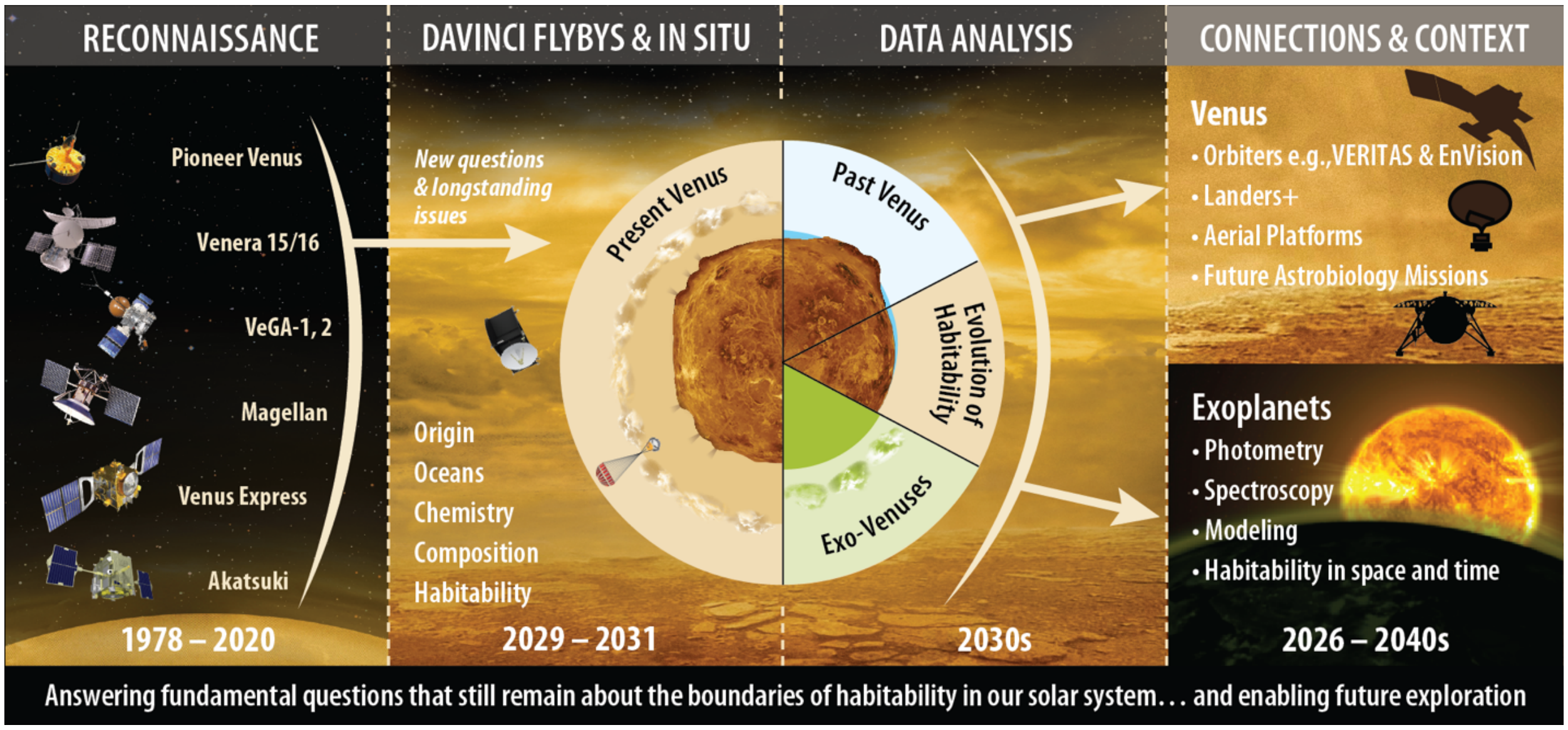

Figure 1: Context for the DAVINCI mission within a framework of past orbital missions and with connections to future missions. Key science themes are highlighted with connections to questions leftover from past missions, and with new connections to contemporary and future missions and the era of exoplanet science. DAVINCI is viewed as a gateway for Venus as a future astrobiology target in the context of how habitability is both established and lost in our solar system and beyond (e.g. Limaye et al. 2021).

Previous Venus exploration has led to significant advancements in our understanding of the bulk atmospheric composition of the planet, its geological history, and its geodynamics (Grinspoon & Bullock 2007; Taylor and Grinspoon 2009; Kane at al. 2019; Way & Del Genio 2019; Lammer et al. 2020). Yet Venus remains the least understood of the inner planets.

With the recent selection of multiple Venus missions, this may soon change. The DAVINCI mission will complement contemporary Venus missions, as shown in Figure 1, which feature next-generation radar and night-side near infrared (NIR) emission spectrometers for mapping the surface at scales from tens of meters (synthetic aperture radar; SAR) to ~100 km (NIR spectroscopy). These payloads will fly on missions in the late 2020s (NASA's VERITAS) and mid-2030s (ESA's *EnVision*) to determine compositional patterns at regional to global scale for advancing models of Venus's crustal and thermal evolution (Ghail et al. 2018; 2021). In turn, the DAVINCI mission will provide *in situ* context for these global remote sensing missions by capturing definitive measurements of atmospheric composition, key atmospheric isotope ratios, multi-band descent imaging, and Venus flyby imaging at ultraviolet (UV) and NIR wavelengths to establish new knowledge about the vertically resolved atmosphere and currently poorly understood regions of the surface.

Venus's thick cloud cover and harsh surface environment in the present day obscure the possibility, supported by recent modeling efforts (e.g. Way & Del Genio 2020), that Venus could have been more Earth-like in the past, possibly even for an extended time period (Figure 2). The hypothesis of a past habitable Venus is supported by accretion models which suggest that Venus and Earth would have had similar initial water inventories (Elkins-Tanton, 2011), by evolutionary climate models (Way et al., 2016), and by the surprisingly elevated ratio of deuterium to hydrogen (D/H) in water in its atmosphere, which is at least 120 times that on Earth. This elevated D/H ratio could result from $H_2O$ photolysis following ocean evaporation, with preferential loss of hydrogen to space compared to the twice-heavier deuterium (e.g. Donahue et al. 1982; Kasting 1988; Donahue et al. 1997). However, other models suggest that Venus never condensed oceans (Hamano et al. 2013; Turbet at al. 2021) and that preferential H loss occurred directly from photolysis of a steam atmosphere. Other possible explanations for the elevated D/H ratio include outgassed water within the past 0.5-1 billion years followed by fractionating escape (Grinspoon 1993). An improved understanding of the history of possible past Venusian water requires improved measurements of the D/H ratio: the Pioneer Venus mass spectrometer measured D/H (~0.016, ~100 × the terrestrial value) after its instrument inlet became clogged with droplets of sulfuric acid (Donahue et al., 1982), and did not survey this key parameter from the top of the atmosphere to the near surface. Ground-based measurements have estimated Venus D/H at 0.019 ± 0.006 or 120 ± 40 × the terrestrial value (De Bergh et al. 1991). More recent Venus Express measurements may be inconsistent with Pioneer Venus and Earth-based observations and imply that the D/H ratio may increase markedly with altitude: Bertaux et al. 2007 measured the bulk lower atmosphere $HDO/H_2O$ at ~0.05, while at 70-95 km, the measured value reached ~0.12, implying imply D/H values of ~0.025 in the bulk atmosphere and up to ~0.06 at 70-95 km.

DAVINCI will provide D/H measurements with high precision (~1% in 10 ppmv; 0.2% in 100 ppmv) to resolve the question of altitude distribution and discriminate between different histories of water loss. D/H measurements in the bulk of the troposphere are missing, making DAVINCI's altitude-resolved measurements particularly important. Additionally, D/H precision of 0.2% is sufficient to resolve between D/H evolution scenarios modeled in Grinspoon (1993). At least one D/H sample will be obtained above the clouds to help resolve between competing hypotheses for the surprising vertical gradient measured by Venus Express (e.g. Liang and Yung 2009), and at least 5 samples will be obtained below 50 km, including at least one below 15 km. In addition to these measurements, hundreds of moderate resolution (20%) mass spectrometer measurements of $H_2O$ and HDO will be obtained from below the clouds to surface touchdown.

Estimates of surface composition from DAVINCI may provide additional corroborating evidence for past oceans. On Earth, silica-enriched *felsic* rocks (specifically granites and granitoids) form from interior continent-building processes with involvement of water (as opposed to *mafic* magmas and rocks, e.g. basalt, which form more commonly in water-poor mantle regions) (e.g. Campbell & Taylor 1983; Filiberto 2014). On Venus, emissivity signatures consistent with felsic rocks have been reported in certain highland regions (e.g. Hashimoto et al. 2008; Weller & Kiefer 2020), including the DAVINCI descent site, the Alpha Regio tessera region (Gilmore et al. 2015).

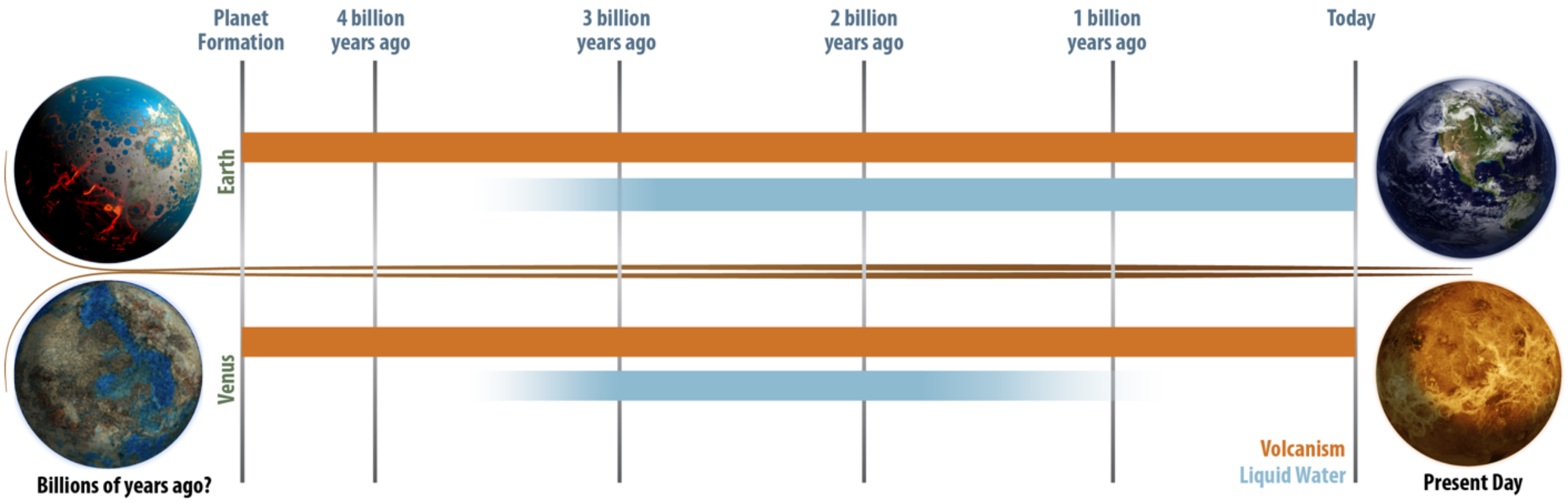

Figure 2. A possible history of water on Venus (e.g. Way et al. (2016); Way and Del Genio (2020)), compared to Earth history. Venus's epoch of surface liquid water may have persisted for over 2 billion years, and DAVINCI measurements can help constrain this hypothesis. Evidence also suggests volcanic activity on Venus persists to this day (e.g. Smrekar et al. 2010), and DAVINCI noble gas measurements will help constrain the history of Venus volcanism.

The evolution of Venus's climate is the result of the interplay between the conditions of formation, the history of solar insolation, the role of exogenous sources of volatiles, and the effects of volcanism over time. DAVINCI measurements of noble gases will provide new insights into all of these processes because, being non-reactive, once released to the atmosphere, they do not react with other material sinks or readily return to the planet's interior. A comparison of noble gases on Venus, Earth, and Mars can provide insights into differences or similarities in the materials that formed each of these planets (e.g., Pepin 2006; Baines et al. 2013; Avice & Marty 2020). The late-1970s measurements from *Pioneer Venus Large Probe* (PVLP) were incomplete and did not offer the precision required to sufficiently measure the noble gases, especially xenon and helium (Lammer et al. 2020). To date, only Ne and Ar have been robustly measured on Venus, rendering it difficult to definitively compare the formation of Venus to Earth and Mars. Neon and argon are both much more abundant on Venus than on Earth --- by factors of 30 and 70, respectively --- and are roughly as abundant as they are in chondritic meteorites (Figure 3). Krypton was measured, but the two reported Kr abundances differ by a factor of fifteen (von Zahn et al. 1983). Only upper limits exist for xenon (Figure 3). The chondritic Ne and Ar abundances suggest a meteoritic source and little subsequent escape. The higher Kr abundance is consistent with this, but the smaller Kr abundance instead suggests a solar nebular source.

The noble gas isotope structures should be more telling. Argon can indicate atmospheric loss through the $^{36}Ar/^{38}Ar$ ratio. Source $^{36}Ar/^{38}Ar$ ratios range from 5.3 (chondritic) to 5.50 (solar), but the $^{36}Ar/^{38}Ar$ is only 4.1±0.1 on Mars (Atreya et al 2013), reflecting a history of atmospheric escape since formation. Neon isotopes can distinguish between nebular and meteoritic sources of the atmosphere. Earth's atmospheric $^{20}Ne/^{22}Ne$ is 9.8 (chondritic), but its interior ratio is 12.5, and rarely greater, possibly reflecting the solar nebula (Williams and Mukhopadhyay 2019). Venus's $^{20}Ne/^{22}Ne$ ratio was reported as 11.8±1.7 or 14±3 by two different missions (von Zahn et al. 1983), which together suggest a protoplanetary nebular source for Ne (with solar ratio is 13.9±0.1; Meshik et al. 2012) rather than a chondritic source with ratio ~10. Confirmation of a solar ratio would be telling. Note, however, that *Viking* reported Mars's $^{36}Ar/^{38}Ar$ ratio as 5.5±1.5, which although correct is also misleading, because the actual ratio hit the bottom of the error bar at 4.1±0.1 (Atreya et al. 2013). It is difficult for Kr to escape from Venus by any process other than impact erosion; hence, it is expected that Kr isotopes should preserve the fingerprints of the source (chondritic for Earth, solar for Mars). Any deviation from this (e.g., a strong mass fractionation) would be revolutionary. Xenon isotopes are the most numerous and have the most potential to see deep into Venus's history. Xenon on Earth and Mars is depleted (i.e., the Kr/Xe ratios are high) and mass fractionated (i.e, the heavy isotopes are relatively more abundant), the latter in particular indicating that, despite its great weight, Xe has escaped. It is hypothesized that Xe escaped as an ion in a photo-ionized magnetically channeled hydrogen wind (Zahnle et al. 2019). In the standard model, Venus lost a great deal of hydrogen to space, probably fairly early in its history, so that early parallel evolution of Earth and Venus might imply that Venus has little or no Xe left. Limited mass fractionation might imply the absence of a planetary magnetic field. A similar isotopic signature in non-radiogenic Xe on Venus and Earth would imply similar energetic processing might have occurred on both planets.

Radiogenic noble gas isotopes produced by decay of parent radionucleotides in the planetary interior, $^{40}$Ar, $^{4}$He, $^{129}$Xe, and $^{136}$Xe, will provide constraints on volcanic outgassing through time (Figure 2). Measurements of $^{40}$Ar are diagnostic of the long-term integrated volcanic outgassing rate (Namiki & Solomon 1998; O'Rourke & Korenaga 2015). In concert with $^{40}$Ar, $^{4}$He provides constraints on the history of volcanic degassing and escape. Meanwhile, $^{129}$Xe and $^{136}$Xe help determine the early and long-term outgassing rate, and also provide information on early impact events. Because the parent of $^{129}$Xe, $^{129}$I, has a 15.7 Myr half-life, the $^{129}$Xe abundance is sensitive to the timing of events during accretion. If $^{129}$Xe is abundant, it would indicate that Venus did not suffer a late giant impact resembling the Moon-forming impact on Earth. Because the parent of $^{136}$Xe, $^{244}$Pu, has an 80 Myr half-life, $^{136}$Xe is more sensitive to events during the first few hundred million years. Information from these noble gases that reveal the timing and history of outgassing on Venus will help test different resurfacing models suggested to explain Venus's 0.2-1 billion-year average surface age based on its ~950 randomly distributed craters (Schaber et al. 1992; Phillips et al. 1992; McKinnon et al. 1997; Herrick & Rumpf 2011; Bottke et al. 2016).

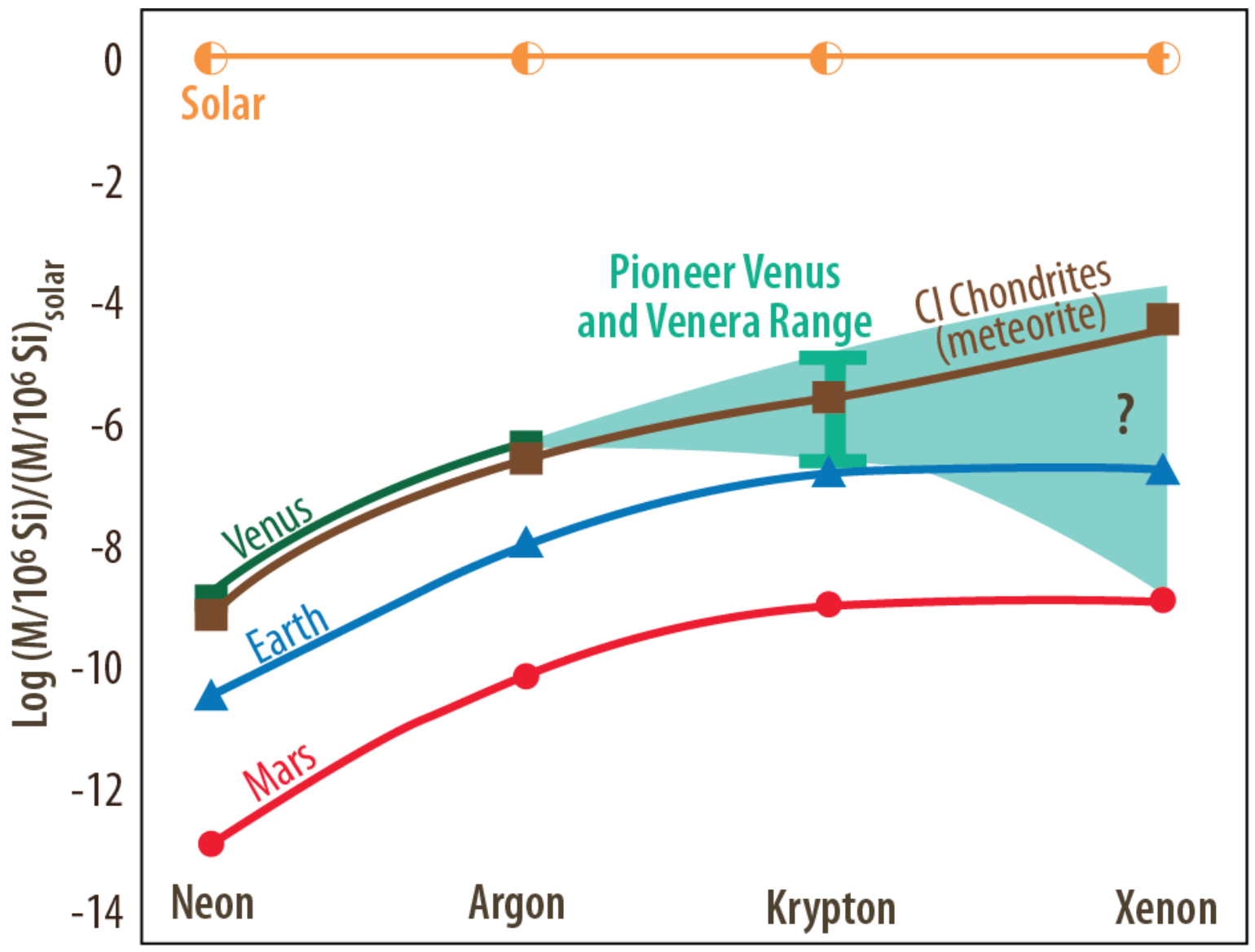

Figure 3. DAVINCI measurements of Kr and its first ever measurements of Xe will resolve questions of differences in the noble gas inventories at Earth (blue curve with triangles), Mars (red curve with circles), and Venus (teal curve with squares, with the Pioneer Venus and Venera range indicated as the filled teal region). For comparison, solar values (orange line with half-circles) and carbonaceous chondrite values (brown curve with squares) are also shown.  These differences may imply variation in materials that formed each planet as well as subsequent events during planetary evolution. Figure after Baines et al. (2013).

DAVINCI measurements of chemically active gases will constrain coupled chemical processes and circulation of the sub-cloud atmosphere. The majority of the atmospheric mass (~75%) on Venus is contained below 20 km where the gas composition is poorly constrained. Compounding our uncertainties, the lapse rate (temperature as a function of altitude) is insufficiently constrained and represents a key variable for current models of the deep atmosphere, where dominant $CO_2$ is super-critical (Lebonnois & Schubert 2017). Vertical composition profiles and gradients in the deep atmosphere are needed to constrain abundances of atmospheric volatiles, physical processes (e.g., circulation, predicted $CO_2$-$N_2$ gas separation, as described in Lebonnois & Schubert, 2017), thermochemical and some photochemical reactions among gases (e.g., Krasnopolsky, 2007, 2013), and the chemical interactions at the atmosphere-surface interface (e.g., Fegley et al. 1997a; Zolotov 2018; 2019). Current data on gas composition above ~10-20 km do not indicate chemical equilibration between gases, except possibly a thin near-surface layer (e.g. Krasnopolsky & Pollack 1994; Fegley et al., 1997b; Krasnopolsky, 2007). Concentration gradients have been observed (CO, OCS, $H_2SO_4$, $SO_3$) or suspected ($H_2O$, $SO_2$) for some gases (e.g. Bertaux et al. 1996; Mills et al. 2007; Marcq et al. 2018). Observed latitudinal anti-correlated of CO and OCS at 33-36 km (e.g. Marcq et al., 2008; Tsang et al., 2008; Arney et al. 2014), indicates latitudinal and altitudinal gradients as well as chemical transformation of the gases to each other (e.g., Yung et al., 2009), coupled with a global circulation. Elevated temperatures in the deep atmosphere should favor formation of OCS from CO and S-bearing gases (Krasnopolsky & Pollack 1994), that suggests increasing OCS abundance towards the surface together with decrease in CO content below ~40 km (Krasnopolsky 2007, 2013; Yung et al. 2009).

Chemically active gases could react with surface minerals and glasses leading to formation of newly formed solids such ferrous compounds that undergo oxidation by atmospheric $CO_2$ and in sulfates and sulfides that trap S-bearing gases (e.g., Fegley et al., 1997a; Zolotov, 2018). In addition to compositional changes, these interactions influence such physical properties of surface materials as grain size, density, electrical conductivity, and reflectance that all affect detectability of altered materials by remote sensing methods (Gilmore et al., 2017). Better understanding of these gas-solid type interactions will require chemical and physical knowledge of the lowest 12 km of the Venus atmosphere. DAVINCI measurements of $H_2O$, $SO_2$, OCS, CO, $H_2S$, sulfur allotropes ($S_n$), and HCl together with temperature-pressure conditions in the deep atmosphere will constrain the stability of primary and secondary solids and inform the directions of gas-solid type reactions. In particular, redox conditions at the atmosphere-surface interface remain uncertain, with fugacity (*f*) of $O_2$ uncertain within almost two orders of magnitude ($\log_{10}fO_2 = 10^{-21.7}$ to $10^{-20.0}$ bars, Fegley et al., 1997b). In the deepest atmosphere, the redox state will be constrained with DAVINCI measurements of major chemically active gases ($CO_2$, $SO_2$, CO, OCS) and $fO_2$ itself will be directly measured with a DAVINCI student collaboration experiment. These measurements will also help determine whether the atmosphere is close to the conditions conducive to varied gas-mineral equilibria (e.g. magnetite-hematite, magnetite-hematite-pyrite), which could assess potential control (buffering) of concentrations of some atmospheric oxidants by surface mineralogy (Figure 4).

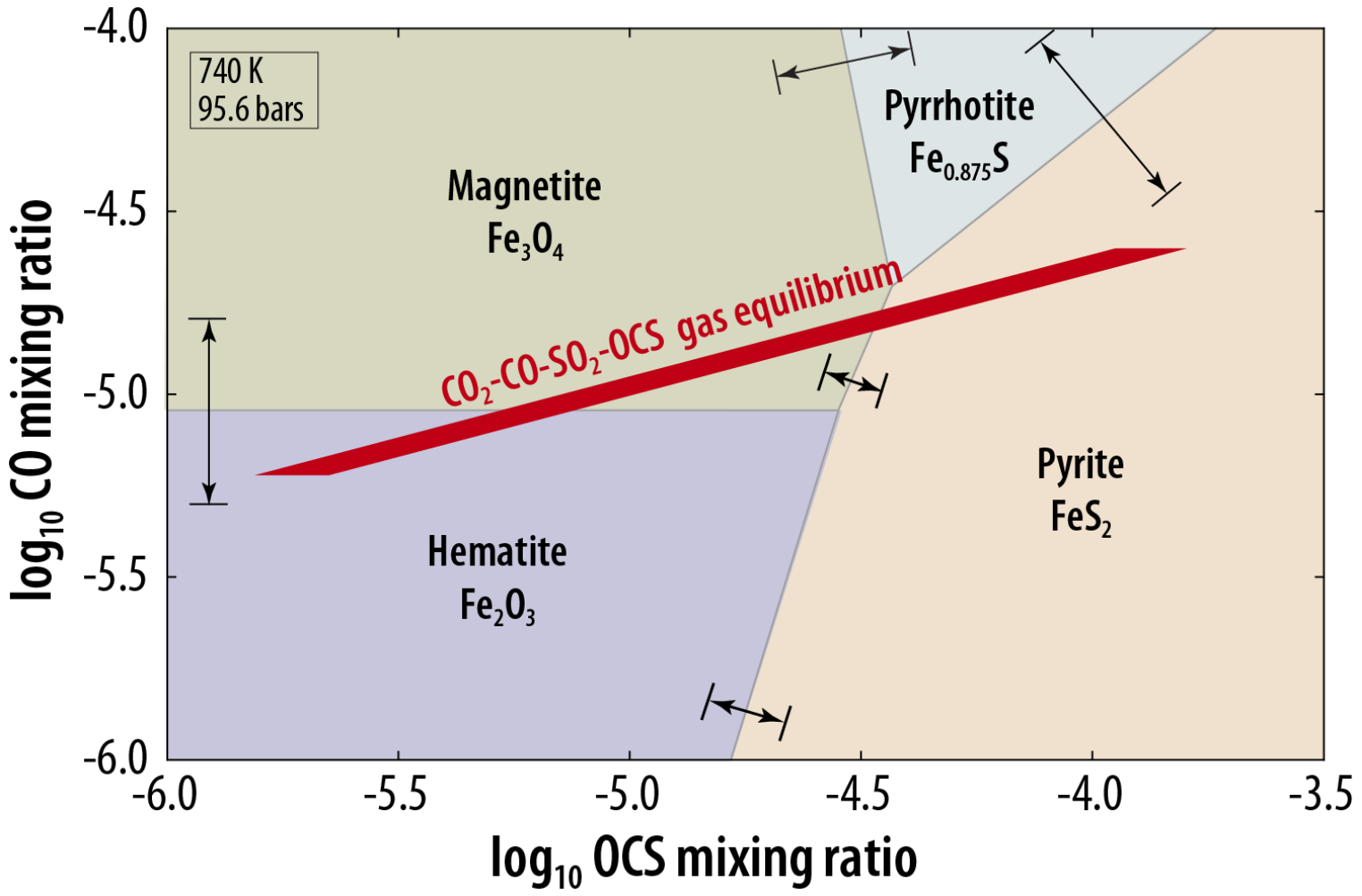

Figure 4. DAVINCI measurements of gas mixing ratios could constrain surface mineralogy and chemical state of the near-surface atmosphere. This diagram shows stability fields of iron-bearing minerals at Venus surface conditions. The red quadrangle corresponds to putative gas chemical equilibrium at mixing ratios of $CO_2$ and $SO_2$ of 0.965 and 130-185 ppm, respectively. Measuring CO, OCS, $SO_2$ and $CO_2$ will determine whether atmospheric gases equilibrate with each other and what minerals are stable. The figure is modified from Zolotov (2015).

Key science questions addressed during the DAVINCI descent are shown in the timeline of Figure 5. DAVINCI fulfills the need for new investigations of the bulk atmosphere by performing measurements of the complete suite of noble gases and confirmation of the D/H ratio in water that together constrain the history of outgassing and atmospheric loss. DAVINCI will conduct definitive in situ analyses of near-surface gases to reveal chemical exchange between the surface and deep atmosphere, and link these *in situ* investigations to new observations of the topography and near infrared reflectivity of a representative tessera to test hypotheses of water-rock interactions that could have led to aqueous minerals, layered water-deposited sediments, and light-colored felsic igneous rocks. Furthermore, the instruments can provide critical compositional context for potential newly discovered species (e.g. $PH_3$; Greaves et al. 2020) that may be linked to the history of habitability on Venus even today (e.g. Limaye et al. 2021), or possibly to ongoing volcanic activity (Truong & Lunine 2021). DAVINCI will also provide a detailed survey of compounds bearing elements critical to life on Earth (e.g., those containing such elements as carbon, hydrogen, nitrogen, oxygen, phosphorus, and sulfur). DAVINCI has been designed to provide flexibility and responsiveness to new discoveries about the Venus atmosphere and will provide vital constraints on key chemical cycles such as those involving sulfur as an example (Figure 6). Because Venus-like exoplanets may represent the most readily observable class of terrestrial worlds for the James Webb Space Telescope (Kane et al. 2014; 2019), measurements at Venus may provide ground-truth to guide and constrain interpretations of these distant worlds as discussed in Section 3 (Gillon et al. 2017; Lincowski et al. 2018; Arney & Kane 2020).

***DAVINCI Mission Science Objectives***

In summary, through its comprehensive suite of measurements, DAVINCI will provide answers to the many scientific questions of our neighboring planet via measurements to be completed during the late 2020s and early 2030s:

1. Atmospheric origin and planetary evolution: What is the origin of Venus’s atmosphere, and how has it evolved? Was there an early ocean on Venus, and if so, when and where did it go? How and why is Venus different than (or similar to) Earth, Mars, and exo-Venuses?
2. Atmospheric composition and surface interaction: Is there any currently active volcanism and what is rate of volcanic activity? How does the atmosphere interact with the surface? What are the chemical and physical processes in the clouds and sub-cloud atmosphere?
3. Surface properties: Are there any signs of past processes in surface morphology and reflectance? How do tesserae compare with other major highlands and lowlands?

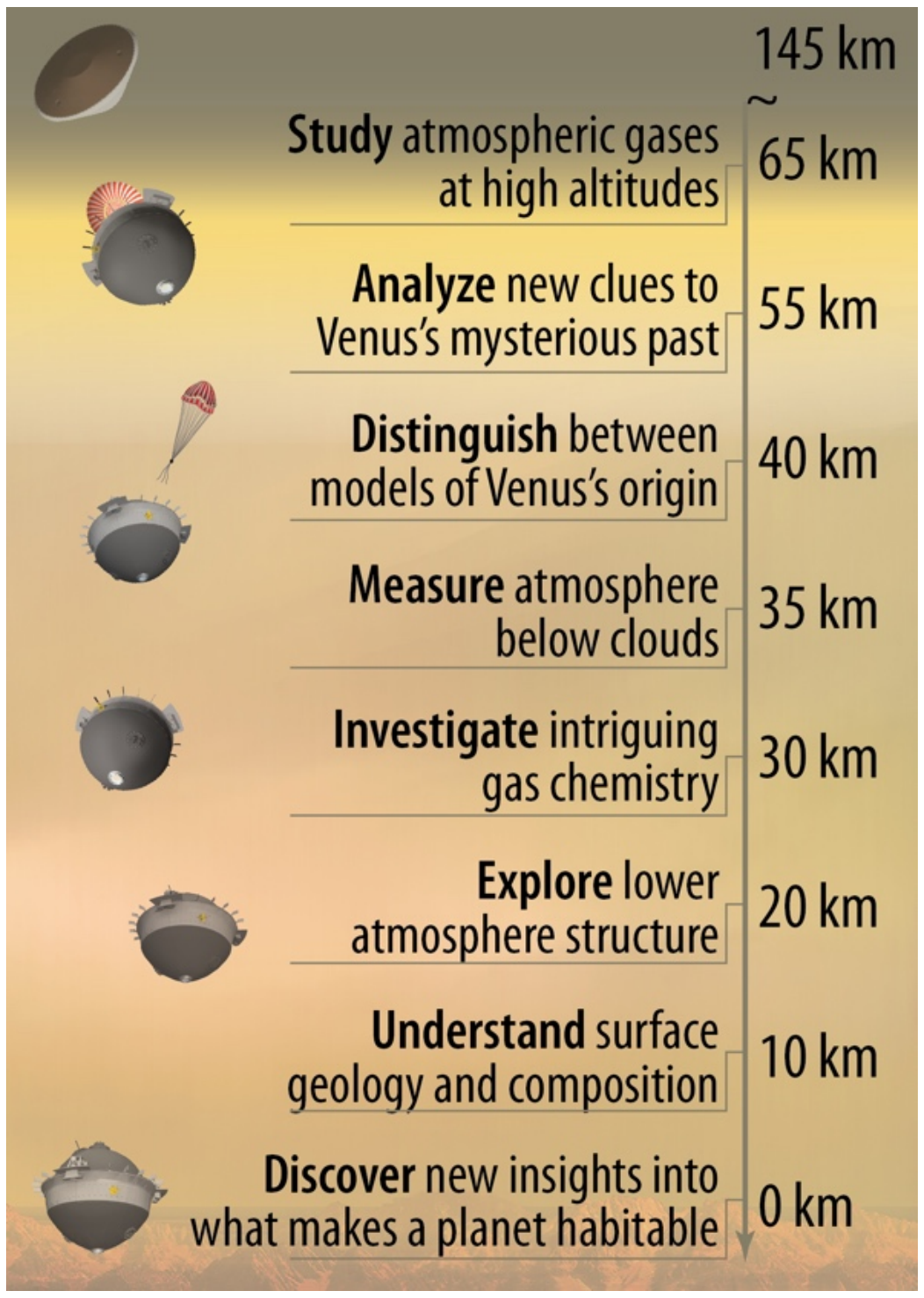

Figure 5: Summary of DS vertical descent timeline in the Venus atmosphere with select science topics pursued at each altitude band.

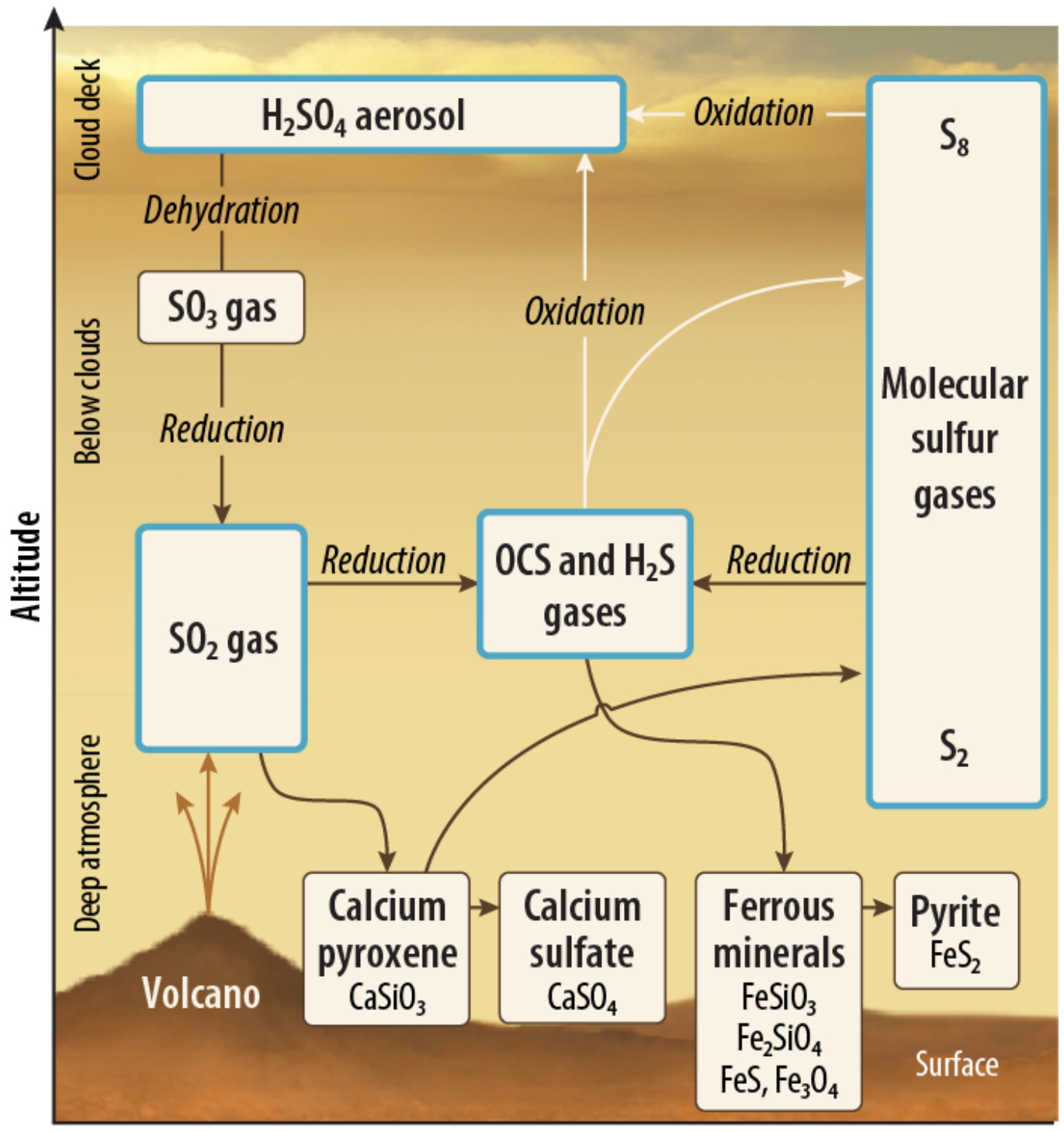

Figure 6: DAVINCI detailed measurements will reveal the composition of the Venus atmosphere below ~70 km, providing necessary context to understand key chemical cycles, such as the putative sulfur (S) cycle shown here. $SO_2$ is the third most abundant gas in the Venus atmosphere after $CO_2$ and $N_2$, so measurements of it and other S-bearing gases are important anchors for Venus atmosphere chemical and physical models. Boxes outlined in blue designate key species targeted by the DAVINCI descent sphere analytical instruments (Section 2.3).

## 2. Mission Design tied to Science Drivers

DAVINCI is a multi-element mission concept that delivers both a deep atmosphere descent sphere (DS) (i.e., a "probe") and a flyby remote sensing Carrier-Relay-Imaging-Spacecraft (CRIS) to Venus, each carrying sophisticated instruments tailored to the prioritized scientific goals and objectives of the mission. As selected by NASA in June 2021, the primary mission design for DAVINCI features two flybys and an *in situ* descent phase that would deliver definitive chemical and isotopic composition of the Venus atmosphere during a 59-minute transect from ~ 70 km to the surface (Figure 5). This *in situ* investigation is preceded by remote observations of the dynamic atmosphere, cloud deck, and surface properties during the flybys, prior to the entry-descent-science *in situ* phase involving the DS. As described in Section 1, this "flyby-probe" mission architecture is optimized to produce a set of focused measurements to improve models of Venus's current and past state, its atmospheric and interior evolution, and questions about habitability (e.g. Way & Del Genio 2020; Limaye et al. 2021; Turbet et al. 2021).

### 2.1 Overall Mission Architecture

As selected, DAVINCI would nominally launch in June 2029 as shown in Figure 7, and after a ~six-month cruise, the spacecraft would fly by Venus for unique remote sensing science that includes dayside UV cloud motion videos, hyperspectral UV imaging spectroscopy, and night side NIR surface emissivity mapping. As currently planned, the trajectory returns nine months later for a second flyby in November 2030 with additional dayside UV observations and nightside surface measurements of key highlands (e.g., tesserae and Maat Mons). The flight system returns to Venus seven months later and delivers the *in situ* descent sphere to *Alpha Regio* on June 21, 2031 with favorable solar illumination for descent high-sensitivity NIR imaging under the clouds. DAVINCI's targeted entry-descent-imaging site within *Alpha*

*Regio* has been comprehensively investigated by prior missions and is large enough (nearly twice the size of Texas) such that a precisely controlled descent is not necessary. DAVINCI's touchdown ellipse comfortably fits within this area with large margin, and enables high-resolution descent images to map the local composition-related infrared emissivity and local topography of this unique region. Figure 8 highlights the descent sphere imaging corridor and its landing error ellipse within Alpha Regio using the Arecibo radio-telescope-based pseudo-topography of this tessera region at sub-km scales.

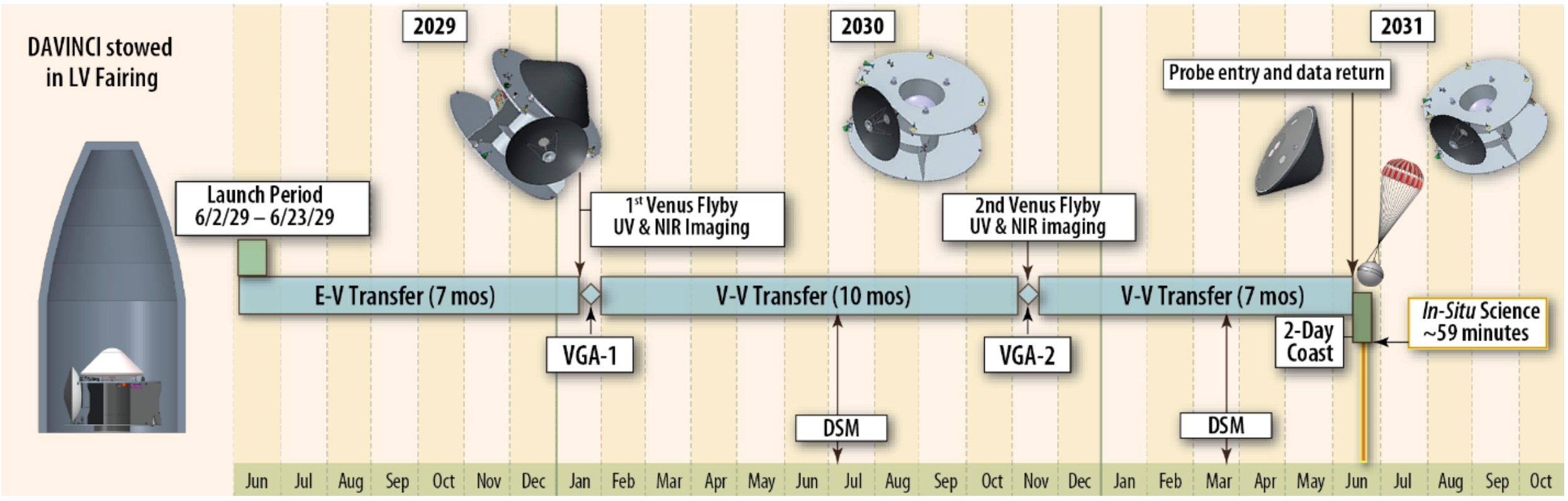

Figure 7: DAVINCI nominal mission timeline from launch in June 2029 through descent sphere release in June 2031. Note that the baseline mission ends in Sept. 2031 after relay of all acquired datasets (descent sphere and flybys) to Earth.

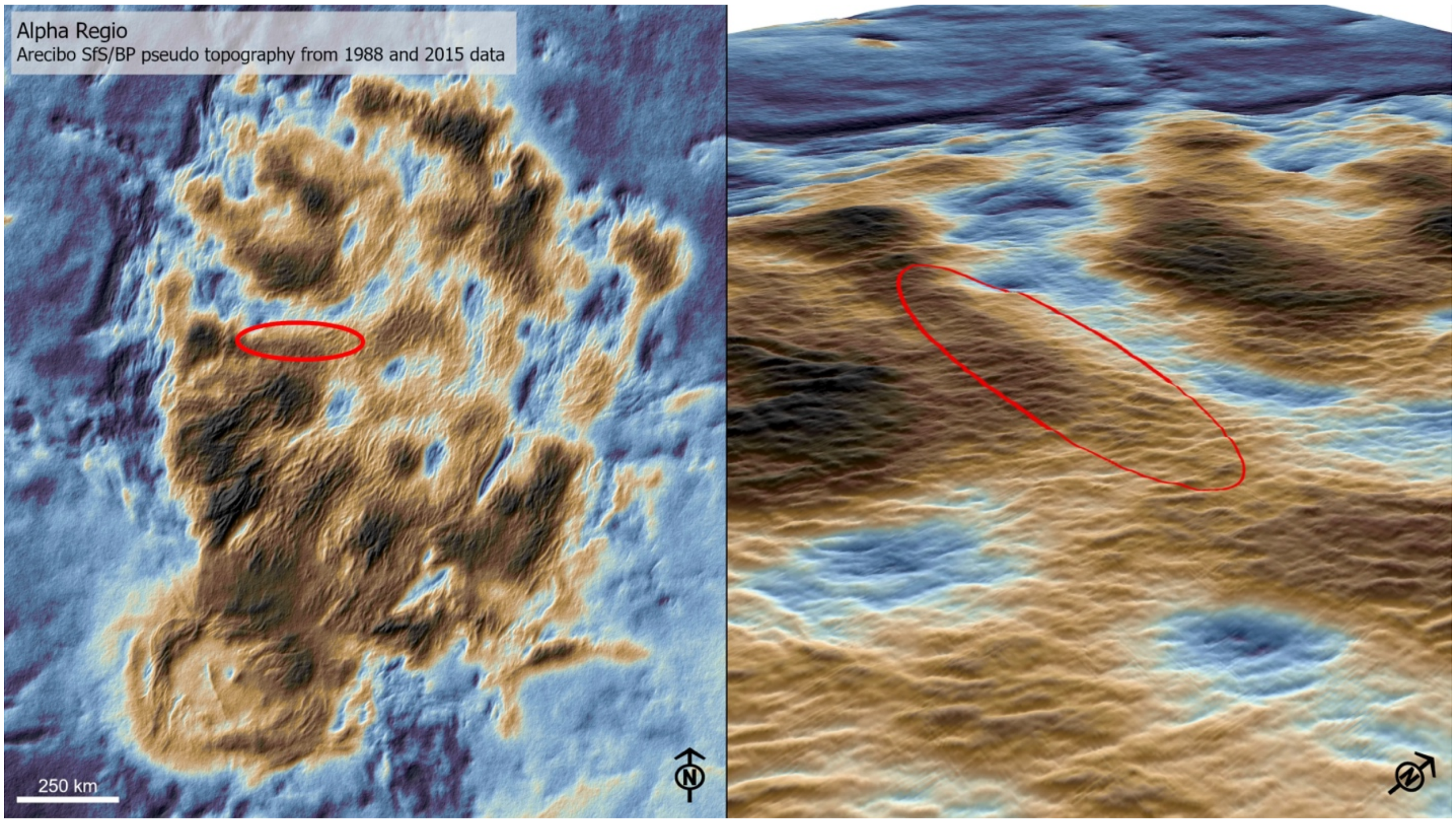

Figure 8: Alpha Regio at approximately 20 S latitude with up to 3 km of total relief above the adjacent plains. Left-side map view is derived from Arecibo Earth-based radar mapping using 1988 and 2015 datasets controlled by Magellan radar altimetry, with the red "ellipse" being the 3-sigma error ellipse that constitutes the imaging descent corridor. The color scaling represents pseudo-topography from low blue (0 km at the mean planetary radius, MPR, of 6051.84 km) to dark brown (over 2.5 km above MPR, AMPR). At right is a perspective view of the entry corridor (red ellipse is ~ 310 km in its long-axis) atop the ridged mountains of Alpha Regio with over 900m of local relief. Arecibo data analysis and processing by the DAVINCI team.

In June 2031, two days before arrival at Venus, the Probe Flight System (PFS) is released. The spacecraft observes its release, and then conducts a divert maneuver to fly by Venus and communicate with the DS throughout the *in situ* science mission within the Venus atmosphere. After atmospheric entry and parachute deployment (~70 km altitude), the heat shield is released and the DS instruments begin to collect and return altitude-resolved high fidelity measurements of noble, trace gas, and isotopic abundances; atmospheric temperature, pressure, and winds; and high-resolution broadband (740-1040 nm) and narrowband (980-1030 nm) NIR descent images of Alpha Regio. Although not required to land on the Venus surface, the DS has sufficient resources to continue conducting science and relaying data for an additional ~18 minutes from the surface if it survives the 18.7 m/s surface touchdown. After CRIS has recorded the required descent sphere data, it turns toward Earth and transmits those data to the Deep Space Network (DSN) via its X-band medium-gain antenna.

Principal DAVINCI mission flight systems are shown in Figure 9. The Lockheed Martin (LM) spacecraft (CRIS) has high heritage from prior planetary missions with NASA's Jet Propulsion Laboratory (JPL) and Goddard Space Flight Center (GSFC). The PFS includes the integrated DS and Entry System (ES). LM will integrate the DS and ES at its facility near Denver, CO. The ES consists of a 45° half-angle sphere-cone entry vehicle consisting of a Carbon-Carbon thermal protection system, heat shield, drogue and main parachutes, and a back-shell assembly. LM is responsible for the ES, and it has heritage from *Genesis* and *Stardust* with additional design aspects from *Pioneer-Venus Large Probe* (PVLP), Mars *Phoenix* polar lander, and the *Mars Science Laboratory* (MSL) rover. The ES encapsulates and protects the DS during its initial Venus atmospheric entry. The DS is a pressure vessel and aero-fairing with five science instruments (Figure 9; Section 2.3). The DS benefits from PVLP flight heritage and extensive GSFC prototype, Engineering Test Unit (ETU) and test efforts at relevant Venus conditions. The science instruments carried aboard the DS and their instrument heritages are:

- Venus Mass Spectrometer (VMS) - Leverages recent successful mass spectrometer designs, including Mars Science Laboratory Sample Analysis at Mars (MSL/SAM) Quadrupole Mass Spectrometer (QMS) (Mahaffy et al. 2012)
- Venus Tunable Laser Spectrometer (VTLS) - Draws flight heritage from MSL/SAM Tunable Laser Spectrometer (TLS)
- Venus Atmospheric Structure Investigation (VASI) - Design heritage from previous atmospheric entry probes for measuring pressure, temperature, and acceleration
- Venus Descent Imager (VenDI) - Heritage from MSL MastCam/MARDI and OSIRIS-Rex NavCams (Ravine et al. 2014; 2016) with large-pixel CCD detector for maximal signal to noise
- Venus Oxygen Fugacity Student Collaboration Experiment (VfOx) – A solid-state nernstian ceramic oxygen sensor with heritage from high temperature industrial sensors (e.g. Sato & Wright 1966; Riegel et al. 2002; Akbar et al. 2006)

The CRIS spacecraft carries two science instruments:

- Venus Imaging System for Observational Reconnaissance (VISOR) – Contains flight-proven components from the OSIRIS-Rex TAGCAMS navigation cameras
- Compact Ultraviolet to Visible Imaging Spectrometer (CUVIS) – A technology demonstration that features new freeform mirror technology and artificial intelligence/machine learning capabilities to enable new ultraviolet hyperspectral and high spectral resolution spectroscopy.

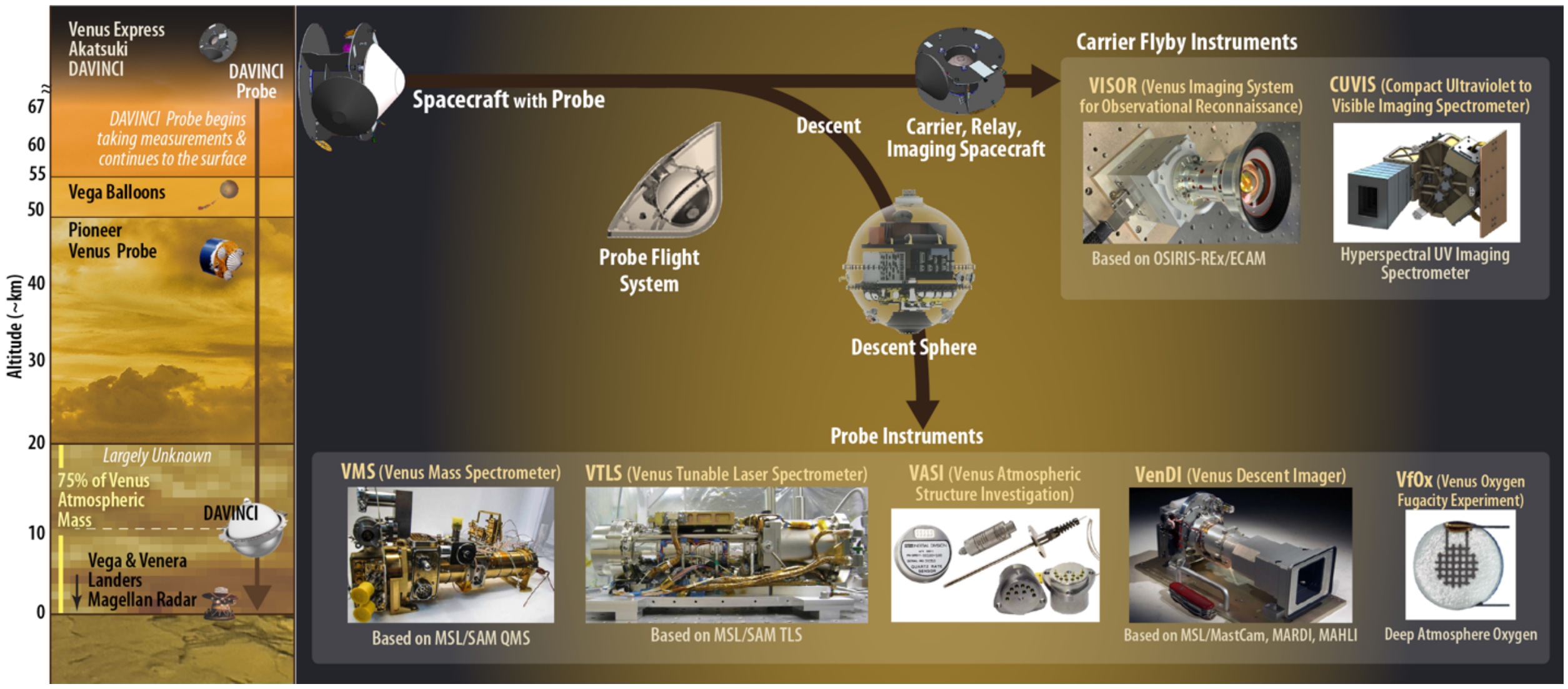

Figure 9: DAVINCI delivers high-priority science using five descent sphere-based (DS) instruments, and two remote sensing instruments on the carrier-relay-imaging spacecraft (CRIS), all with robust flight heritage. See Table 2 for instrument details for the *in situ* descent sphere in comparison with previous in situ missions (PVLP).

## 2.2 Descent Sphere Design

The DAVINCI DS is a hermetically sealed titanium pressure vessel with dimensions (0.98 m x 0.85 m; 200 kg) similar to the *Pioneer Venus Large Probe* (PVLP). It is notable that more recent work in descent sphere design reinforces the importance of maintaining and developing new probe technologies to explore the Venus atmosphere (Lorenz 1998; Hall et al. 2000; Israelavitz & Hall 2020). Inside the DS are two payload decks, one forward and one aft relative to the descent vector. The two decks are joined together with a frustum and attached to the pressure vessel with titanium isolators around the perimeter of the forward deck. The pressure vessel itself consists of two hemispheres, forward and aft, and a mid-ring. Aerodynamic properties of the sphere are trimmed using a titanium faring joined to the forward hemisphere and mid-ring to define the outer mold line, drag plates to limit the descent velocity, and a set of spin vanes around the perimeter of the mid-ring to provide a controlled spin rate as the sphere descends through the Venus atmosphere. The main parachute bridle attaches to points on the mid-ring. Figure 10 illustrates the outer DS components.

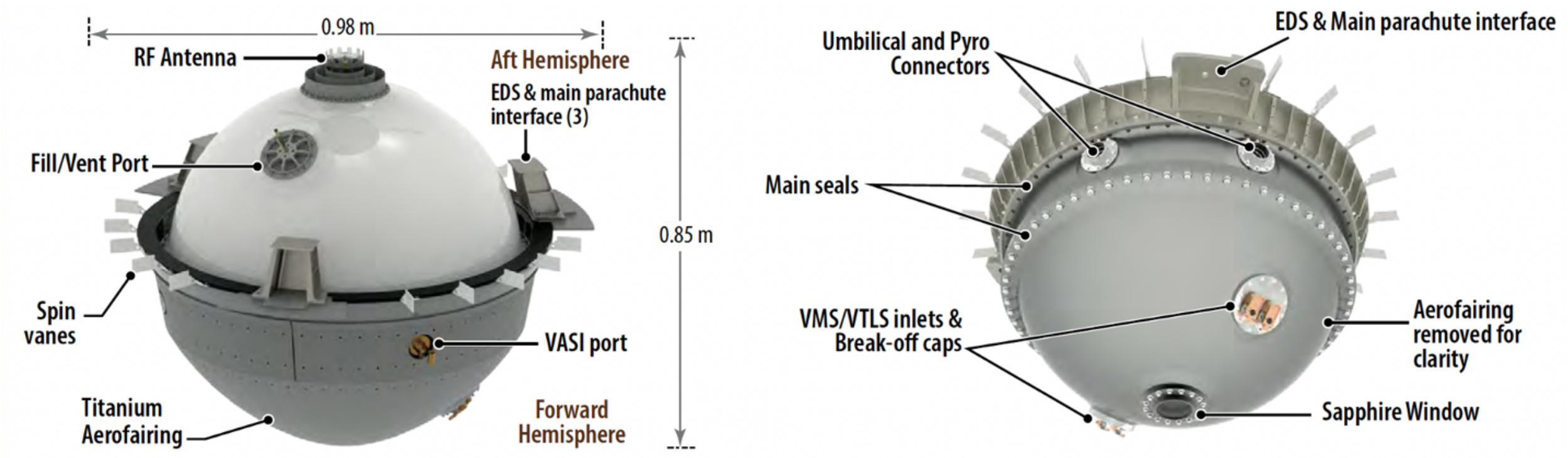

Figure 10. DAVINCI’s descent sphere protects the instruments inside from the harsh Venus ambient environment.

Three sets of inlet ports provide access to the Venus atmosphere for the VASI, VMS, and VTLS instruments (Figure. 9), and a sapphire window oriented at nadir provides a view to the surface for the VenDI camera. Other pressure vessel penetrations provide feedthroughs for connections to the spacecraft and ES during cruise and descent, an omni-directional antenna on the top of the sphere to relay telemetry to the spacecraft, and a fill/vent port for pressurizing the DS prior to launch.

All of the science instruments are mounted on the forward deck. Plumbing connects each atmospheric sensor to its respective port. The VMS and VTLS inlet ports, totaling four inlets, are fitted with break-off caps that are ejected at the appropriate times to allow atmospheric gas ingestion at different altitudes. The aft deck accommodates battery to power the DS after separation from the spacecraft; an adaptive transponder; avionics to execute the descent timeline activities, collect, store and forward science data; internal pressure and acceleration sensors; and a small gas re-pressurization system used in the event of pressure decay during cruise.

To protect against external temperatures that increase during descent and reach up to 460°C at the surface, the temperatures of the internal components of the DS are maintained within their operational limits during the descent through the Venus atmosphere using several passive thermal control techniques refined during GSFC design and test efforts. The DS benefits from over ten years of investment and engineering refinement at GSFC including testing in representative Venus environments (Figure 11; Table 1). Test conditions have not only reproduced the extremes of relevant temperatures and pressures of the Venus surface, but testing has been conducted under ramped temperature and pressure profiles to reproduce day-in-the-life environmental conditions specific to the mission design. A high-emissivity coating on the outer surface of the sphere aids in cold biasing the internal components prior to descent. A combination of insulation types and flexures help protect from radiative and convection effects as well as isolate the decks. In addition, phase change material is utilized around some assemblies that are by necessity near the outer wall. Finally, low-emissivity coatings are used where needed to minimize radiative transfer. All of these measures ensure the extreme environment of Venus does not affect DAVINCI instrument performance.

The ~1 hour descent sequence is shown in summary in Figure 5 and in detail in Figure 12. Monte Carlo simulations of the sequence have been performed over the past five years using current Venus atmosphere reference models, with results that were independently checked against performance requirements throughout the mission proposal review process. The PFS separates from the carrier spacecraft two days before the descent and the SC commences tracking with the High Gain Antenna (HGA) using a two-way S-band link. Entry begins at approximately 145 km altitude at the Atmospheric Entry Interface (AEI) and after a brief blackout period, the pilot chute deploys, the ES back-shell separates, and the main chute is deployed. The ES heatshield is separated and the DS descends while connected to the main parachute. Ingestion ports are open and first VTLS acquires critical samples of the upper atmosphere, then VMS acquires additional samples; both instruments continue to acquire and analyze samples throughout the descent. Approximately 32 minutes after AEI, the main chute separates for the terminal descent and VenDI begins acquiring nadir-oriented NIR images until touchdown with high enough SNR (> 70:1) to observe surface features at meter scales and potentially discern compositional patterns (Garvin et al. 2018; 2020) at broader scales (10-100m). Data are continuously transmitted to the overflying CRIS during the descent through touchdown, but the DS is not required to survive the impact at the time of touchdown, and all science goals are met prior to this.

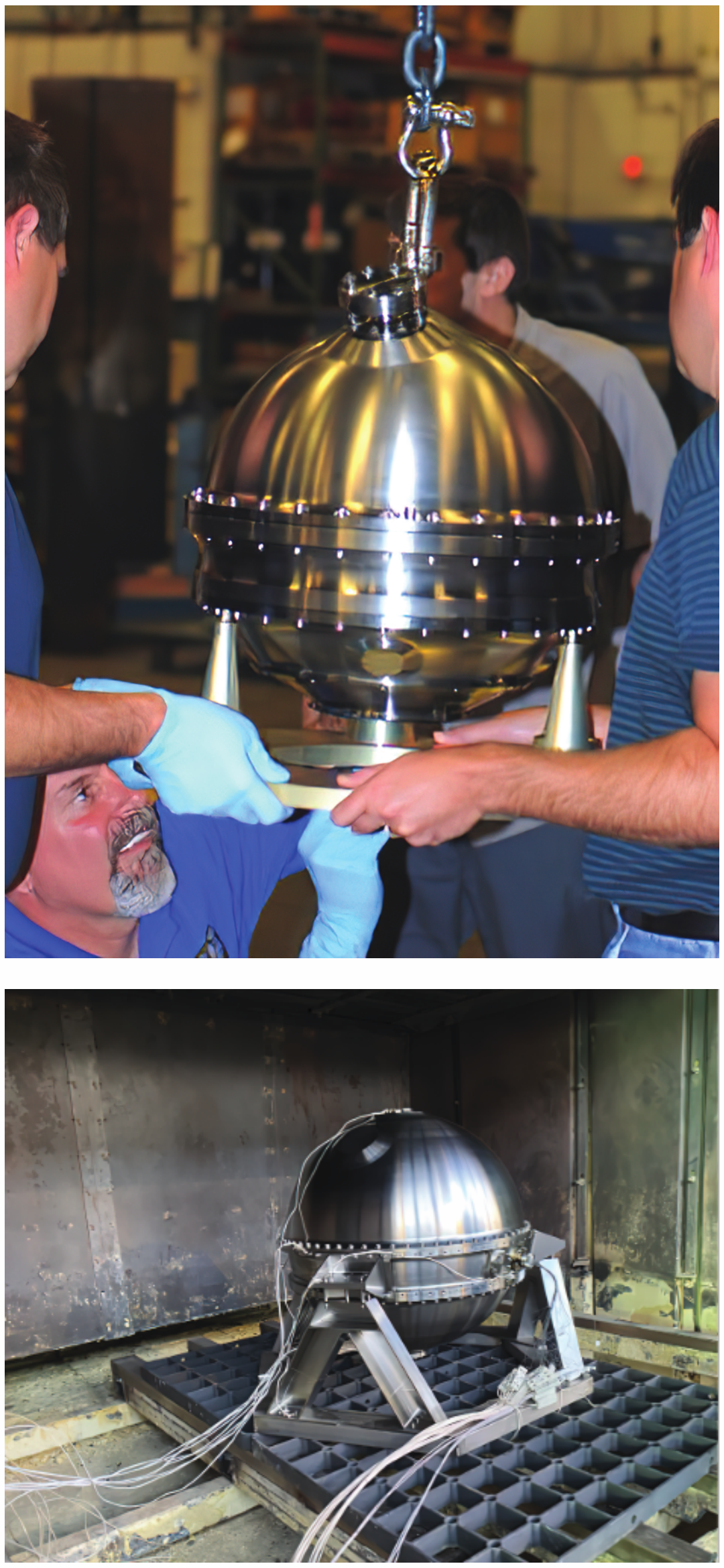

Figure 11: Top panel: A DAVINCI half-scale Engineering Test Unit (ETU) descent sphere before a Venus environment test as part of the development of the overall mission concept. Bottom panel: A full-scale DS ETU was tested to Venus temperature profiles in Jan.-Feb. 2021, with successful verification of performance at temperature. Diameter of the sphere is 0.98 m. Cables are related to engineering testing apparatus. This ETU validated the descent timeline temperature profile performance from ~70 km down to the Venus surface during the DAVINCI mission Phase A activities.

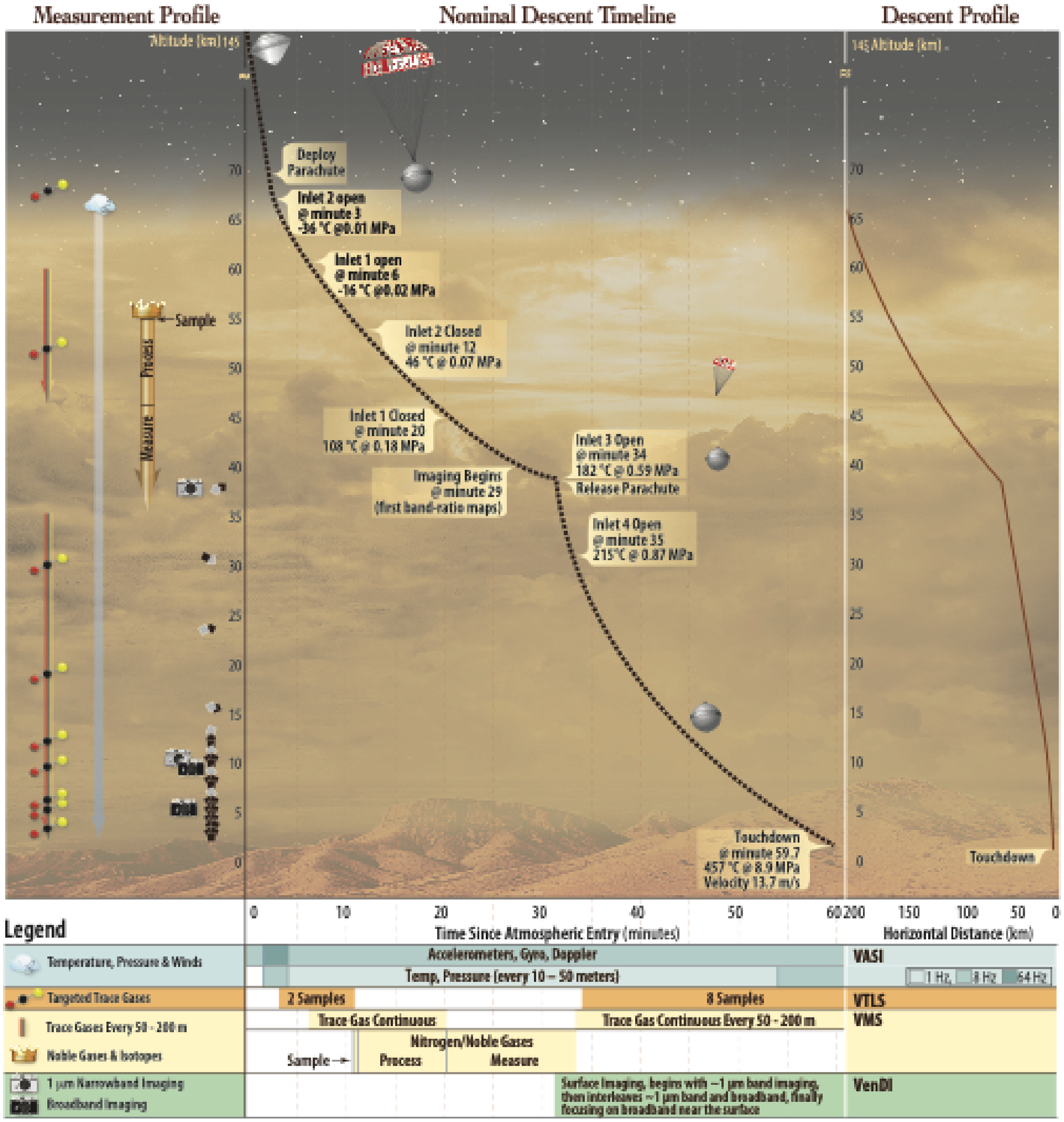

Figure 12: The DAVINCI descent timeline is a carefully choreographed sequence of events. Surface touchdown occurs at 57.04 to 66.7 minutes at 99% confidence. The timeline for the reference touchdown at 59.7 minutes is shown here.

**Table 1: Descent Sphere development testing.**

| Test | Pressure (atm) | Temperature (C) | Details |
|---|---|---|---|
| Hemi-Sphere | 1 | Cold to 460C | Measured thermal blanket performance on unsealed stainless-steel hemisphere |
| Half-Scale descent sphere | 1 to 118 | Ambient to 450 | Successful pathfinder for future DS designs with lessons on seals and connectors. (Figure 11). Several tests were conducted to various conditions, including separate pressure and temperature tests, then final combined pressure and temperature test. |
| Descent Sphere Interfaces | 1 to 95 | 20 to 460 | Successfully tested individual components: inlet ports, metallic seals, umbilical connector, RF connector, fill and vent port |
| Two Piece Ports | 1 to 95 | 20 to 460 | Successfully tested combined components on a test fixture: larger metallic seal, umbilical connector with test harness, fill and vent port, VenDI window |
| VenDI Window | 190 | 490 | Demonstrated window leak rate was within requirements over 4 thermal cycles |
| Full Scale descent sphere | 1 | 20 to 500 | Fabricated full-scale titanium sphere, practiced assembly, and handling. Tested at metal foundry heat-treating facility to reach temperature. Successfully met requirements and correlated thermal model. Temperature testing successful. |

### 2.3 Descent Sphere Payload

The DAVINCI mission will explore Venus and its atmosphere through a carefully architected *in situ* mission rich in comprehensive measurements. The DAVINCI DS utilizes five instruments to bring a highly capable analytical chemistry laboratory (Table 2) that greatly advances beyond the Pioneer Venus Large Probe payload into the Venus atmosphere, in conjunction with a high contrast NIR descent imaging system and an oxygen fugacity sensor to be built as a student collaboration experiment.

Table 2: Comparison of DAVINCI descent sphere-based instruments to those of the 1978 Pioneer Venus Large Probe (PVLP) (Donahue 1982), with specific details listed in right-most column (Bougher et al. 1997, Crisp et al. 2002;).

| DAVINCI Descent Sphere | Pioneer Venus Large Probe (PLVP) | Comparison of DAVINCI to PLVP |
|---|---|---|
| **VMS: Venus Mass Spectrometer** Quadrupole Mass Spectrometer for noble gas measurements in the bulk atmosphere and composition measurements at 50-200 m cadence in altitude | **NMS: Neutral Mass Spectrometer** Magnetic sector mass spectrometer Direct noble gases: Ne, Ar, Kr | DAVINCI offers improved sensitivity to noble gases, from He to Xe, a wide mass range for broad compositional measurements, improved inlet design |
| **VTLS: Venus Tunable Laser Spectrometer** High precision isotope ratios for D/H, C-, O-, and S-bearing species as a function of altitude | **GC: Gas Chromatograph** Discrimination of $N_2$/CO, corroboration of atmospheric chemistry | DAVINCI offers tailored and targeted analytical capabilities to address the need for high precision isotopic measurements |
| **VenDI: Venus Descent Imager** Broadband and narrowband infrared channels for tessera imaging beneath the cloud deck | No comparable geologic study of descent region | DAVINCI offers unprecedented spatial resolution and high sensitivity, with modern data processing methods, to constrain composition and morphology |
| **VASI: Venus Atmospheric Structure Investigation** High cadence measurement of temperature, pressure, wind speed acceleration | **Temperature, pressure, and acceleration sensors** Established structure of atmosphere, with wind measurements, evidence for wave activity in lower | DAVINCI will provide important contextual measurements of atmospheric structure, with first lapse rate measurement in lower |
| | atmosphere | atmosphere |
| **VfOx: Venus Oxygen Fugacity Student Collaboration Experiment** Small ceramic sensor measures oxygen fugacity in the lower atmosphere | No comparable study of oxygen fugacity | DAVINCI will provide sensitive altitude-resolved measurements of atmospheric oxygen in the lower atmosphere |

***Venus Mass Spectrometer (VMS)*:** VMS is a quadrupole mass spectrometer (QMS) with a gas enrichment system and pumping system that will provide the first comprehensive *in situ* survey of the planet's noble gases to reveal Venus's origin and evolution. Leveraging heritage from the Mars Science Laboratory (MSL) Sample Analysis at Mars (SAM) QMS (e.g. Mahaffy et al. 2012; Atreya et al. 2013; Webster & Mahaffy 2011) and with a broad mass range from 2-550 Da, VMS has the capability to discover new trace gas species. VMS acquires hundreds of trace atmospheric constituent mixing ratio measurements and composition measurements during the descent for understanding present-day chemical processes and cycles in the Venus atmosphere (Figure 6). These trace gases measurements are vital for understanding the origin of Venus's atmosphere and its divergent evolution, compared to Earth's. Measurements will occur every ~200 m or better below 61 km, particularly in the lowest 16 km of the atmosphere (Figure 13). Pressures in the sampling lines are controlled with carefully-sized restrictors and capillary leaks. Two independent inlets and sampling lines are used during the descent, providing additional range to accommodate the increasing pressure during the descent. Previously, the Pioneer Venus Large Neutral Probe Mass Spectrometer suffered a clog from a sulfuric acid droplet. To avoid this on DAVINCI, VMS incorporates heated inlet tubes to vaporize trapped droplets, filters of passivated/sintered metal spheres to capture particles large enough to cause clogs in capillary leaks used for pressure reduction, and the aforementioned second inlet for sampling below the sulfuric acid cloud and haze. Table 3 provides a selection of species VMS (and VTLS) will measure, together with current known values and projected accuracy as of current Phase B.

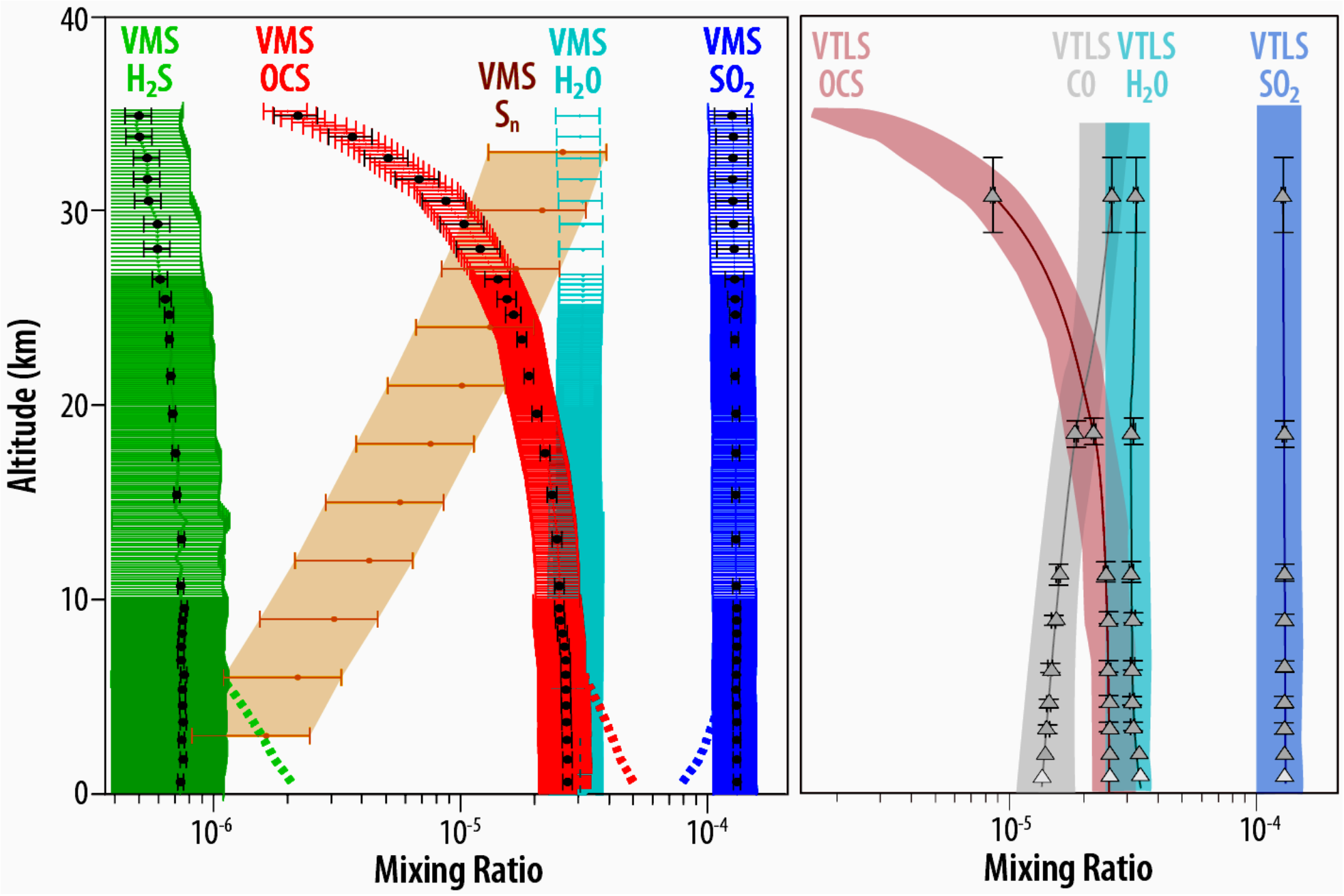

Figure 13. Representative altitude-sampled measurements of selected key species for VMS (high cadence colored points in left panel) and VTLS (lower cadence gray triangular points in right panel) in the lower atmosphere. Averaging of VMS values (left panel) can achieve smaller uncertainties without significant loss of vertical structure information as illustrated with the black points with reduced error bars. Possible gradients near the surface are indicated with dashed lines at altitudes < 10 km in the VMS (left) panel. Additional VTLS measurements beyond the DAVINCI reference mission scenario may be possible, as suggested by the notional white VTLS points in the deepest part of the atmosphere (right panel).

***Venus Tunable Laser Spectrometer (VTLS)*:** VTLS consists of a multi-pass Herriott cell with three laser channels at 2.64, 4.8 and 7.4 μm, specifically targeting key science questions that discriminate chemical processes in the upper clouds and near-surface environment. VTLS draws heritage from the MSL SAM tunable laser spectrometer (e.g. Webster & Mahaffy 2011; Mahaffy et al. 2012; Pla-Garcia et al. 2019). A fourth laser channel is the subject of an ongoing trade study to optimize scientific capability without exceeding the as-designed engineering envelope of the VTLS instrument. VTLS is specifically tailored to answer critical questions about the Venus atmosphere by providing the first highly sensitive in situ measurements of key gas species containing H, S, C and O, as well as their high precision isotope ratios including D/H. VTLS measures gases from at least one sample ingested in the upper cloud, and at least five vertically distributed measurements below the cloud, including one in the lowest 15 km. The exact number of measurements will be determined by operational parameters, such as descent time and data transmission rates during the mission. Selected VTLS lower atmosphere measurements are shown in Figure 13 and described in Table 3.

**Table 3. Selected subset of species measured by VMS and VTLS.**

| **Example Species** | **Value at Venus (*best current knowledge*)** | **Current Uncertainty** | **Altitude Dependent ?** | **DAVINCI projected accuracy (as of Phase B)** | **Reference(s)** |
|---|---|---|---|---|---|
| $H_2O$ | 30 ppm (<45 km) | up to 50% | expected | 20% [VMS], 2% [VTLS] | Taylor et al. 1997; Chamberlain et al. 2013; Arney et al. 2014; |
| D/H in $H_2O$ | 0.016 (~54 km), 0.06 (70-95 km) | 13% | expected | 1% in 10ppmv, 0.2% in 100ppmv [VTLS] | Donahue et al. 1982; deBergh et al. 1991; Bertaux et al. 2007 |
| CO | 20-40 ppm (20-45 km) | up to 60% | expected | 2% [VTLS] | Oyama et al. 1980; Marcq et al. 2006; Cotton et al. 2012 |
| OCS | 0.44-0.55 ppm (36 km), 4.4 ppm (33 km) | up to 29% | expected | 20% [VMS], 2% [VTLS] | Pollack et al. 1993; Taylor et al. 1997; Marcq et al. 2006; Arney et al. 2014 |
| $SO_2$ | 130-150 ppm (< 45 km) | up to 40% | expected | 15% [VMS], 2% [VTLS] | von Zahn et al. 1983; Marcq et al. 2008 |
| $^{32}S/^{33}S/^{34}S$ in $SO_2$, OCS | unknown | unknown | expected | 1‰ [VTLS] | No measured value |
| $H_2S$ | 3 ppm (<24 km) | 67% | expected | few ppm (best effort) [VMS] | Hoffman et al. 1980 |
| $H_2SO_4$ | 8 ppm (~46 km) | unknown | expected | few ppm (best effort) [VMS] | Jenkins et al. 2002 |
| $S_n$ | ppb expected | unknown | expected | few ppb (best effort) | No measured value |
| | | | | [VMS] | |
| He | 9 ppm (~100 km) | 67% | not expected | <4% [VMS] | Krasnopolsky and Gladstone (2005) |
| Ne | 7 ppm | 43% | not expected | <5% [VMS] | von Zahn et al. 1983 |
| Ar | 70 ppm | 36% | not expected | <2% [VMS] | von Zahn et al. 1983 |
| Kr | 50-700 ppm | up to 50% | not expected | <5% [VMS] | von Zahn et al. 1983 |
| Xe | unknown | unknown | not expected | <5% [VMS] | No measured value |

***Venus Atmospheric Structure Investigation (VASI)*:** VASI is a suite of sensors that measure atmospheric pressure, temperature, and dynamics. Dynamics will be measured from DS motions in the Venus atmosphere through entry and descent. These data are used to reconstruct the descent profile and to provide thermodynamic context for each atmospheric sample ingested by VMS and VTLS. Internally mounted accelerometers and gyroscopes combined with Doppler tracking via the Spacecraft-to-DS communications link enables detailed reconstruction of the DS path from the top of the atmosphere to the surface and measurement of the vehicle dynamics in support of NASA's Engineering Science Investigation (ESI) to feed forward into the design of future missions. Temperature and pressure measurements via sensors and Kiel probes on externally mounted booms enable *in situ* environmental measurements during the descent. VASI aims to determine the temperature profile to better than 1 K to constrain models and to permit improved calibration of emissivity retrievals, which depend on knowing the temperature of the Venus surface. In addition to their high quality, atmospheric structure data will be obtained with much higher vertical resolution (<50 m) than previous missions.

***Venus Descent Imager (VenDI)*:** VenDI is a NIR descent imaging system with a nadir orientation and 1024x1024 pixel full-frame CCD detector permitting high SNR imaging from under the clouds and sub-cloud haze (~38 km) to the surface of Venus at spatial scales from 1-200 m. The VenDI camera head is based on the heritage design from MSL/MastCam, MSL/MAHLI, and MSL/MARDI (e.g. Malin et al. 2017). Its broadband (740-1040 nm) and narrow-band filters (980-1030 nm) will provide images at spatial scales (< 200 m down to 2 m/pixel) not possible from orbit. These data will be used to constrain surface composition (i.e., distinguish rocks that are felsic from ones that are mafic) by utilizing band ratios, a technique used effectively with data from various sensors and platforms on many planetary surfaces (e.g., Robinson et al., 2007; Delamere et al., 2010; Gilmore et al., 2008). VenDI will acquire bundles of images from which topography can be derived using machine vision algorithms via Structure-from-Motion (SfM), a method that employs multiple overlapping images to infer three-dimensional texture (Garvin et al. 2018). Topography with meter-scale vertical precision can be computed from bundles of VenDI descent images acquired in the lower-most 5 km of descent, with horizontal (spatial) resolution of 10m and finer. Images from ~1.5 km to the surface will feature spatial resolution less than 1-meter allowing erosional studies relating to the environmental history of Venus. Final VenDI imaging resolutions is expected to be < 50 cm/pixel and as fine as 10 cm/pixel depending on two-way data links between the DS and the overhead CRIS in the last moments before touchdown in June 2031. Figure 14 shows an example digital elevation map (DEM) and overlaid band ratio map of the Zagros Mountains on Earth, a terrain with comparable topography to Alpha Regio.

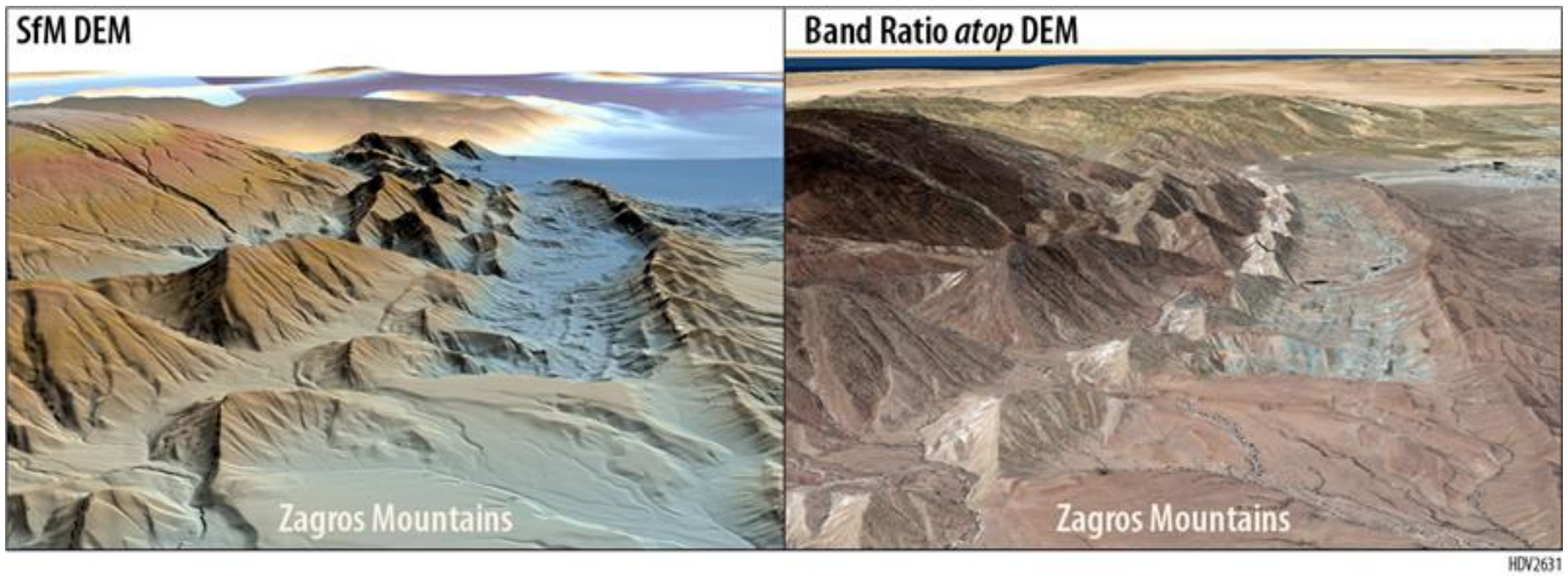

Figure 14. VenDI simulation utilizing data from the Zagros Mountains in Iran as a tessera analog at scales anticipated during the DS descent (FOV 7 km x 7km). Satellite image data courtesy Maxar WorldView (WV-02) processed by NASA Goddard to produce VenDI-like band-ratio maps and to construct a ~ 3m ground scale distance digital elevation map (DEM), as shown here in a color-scaled perspective view. DAVINCI's VenDI will produce similar datasets for Alpha Regio from altitudes below ~ 7 km, depending on final descent timeline during the actual DS entry-descent-science phase. At left: The SfM-based DEM has been ray-traced to be an oblique view to highlight geological structures at 30 m scale for stratigraphic analysis. At right: Band ratio compositional imaging overlain on the DEM will allow potential identification of felsic rocks on Venus in their stratigraphic settings.

***Venus Oxygen Fugacity Student Collaboration Experiment (VfOx)***: VƒOx is a solid-state nernstian ceramic oxygen sensor that relies on a reference material with known oxygen fugacity, $fO_2$ (e.g., a gas mixture or solid oxide). The $fO_2$ differential between the known and unknown sample causes a diffusion of oxygen through the electrolyte, resulting in a small, measurable voltage. VfOx will measure oxygen composition of the lower atmosphere of Venus, with a particular emphasis on informing the oxidation state of surface rocks at our descent location and providing constraints on surface-atmosphere exchange chemistry.

## 2.4 Carrier-Relay-Imaging Spacecraft (CRIS) Flyby Remote Sensing Payload

The DAVINCI CRIS flyby remote sensing payload consists of two instrument packages: (1) Venus Imaging System for Observational Reconnaissance (VISOR); and (2) Compact Ultraviolet to Visible Imaging Spectrometer (CUVIS).

***Venus Imaging System for Observational Reconnaissance (VISOR):*** VISOR is an integrated system of four cameras and controller unit that provides global dayside coverage of Venus in the ultraviolet and nightside coverage in the NIR (0.93 to 1.03 μm) and video of the PFS deployment, all with limited resource requirements and with high heritage. VISOR is based on the Malin Space Science Systems (MSSS) Engineering Camera (ECAM) system (Ravine et al. 2014; 2016), a modular spaceflight imaging system that is currently flying on the OSIRIS-Rex mission (TAGCams) to asteroid *Bennu* and several other missions. Each of the VISOR cameras has a field of view of 11.3° by 8.9° and a format of 2592 by 2048 pixels, which can be converted to a spatial sampling scale (resolution) as a function of range to target. One of the VISOR cameras provides global, dayside coverage of Venus in the unknown UV absorber band (355 to 375 nm). During the flybys, the field of view of this UV camera will cover the full disk of the sunlit planet. The scale of these images will range from 10 to 20 km/pixel at 80,000 to 200,000 km altitude. The other three VISOR cameras image Venus in three independent NIR bands, from 930 to 938 nm, 947 to 964 nm, and 990 to 1030 nm. These near IR bands are used to correct scattered light, correct for variations in cloud layer opacity, and image thermal emission from highland targets on the nightside of Venus (during the two Venus flybys), respectively, to constrain variations of surface emissivity and its correlation with surface geology at regional scales (~100 km, with the spatial resolution limited by the scattering footprint of the Venus atmosphere).

VISOR targets include the DAVINCI descent sphere landing site in *Alpha Regio,* which will allow comparison of the VISOR results for *Alpha Regio* with those acquired during descent sphere descent by VenDI from under the cloud deck. In addition, VISOR nightside NIR imaging will target other highlands to enable comparisons with *Alpha Regio*.

***The Compact Ultraviolet to Visible Imaging Spectrometer (CUVIS):*** This technology demonstration option combines high-resolution UV spectroscopy and hyperspectral imaging from the UV to the visible in a compact package made possible by novel freeform optics and artificial intelligence/machine learning (AI/ML) on-board data processing. A machine learning algorithm based on Generative Adversarial Network (Goodfellow et al. 2014) will be employed for atmospheric parameter retrievals. This will demonstrate how complex tasks can be performed by an AI-enabled device in the on-board data handling system to analyze data on-board in near real time, generate a reduced dataset to be returned in full, and to help flag and prioritize full resolution data to return. With these new capabilities, CUVIS will obtain spectra that are far better for diagnosing upper cloud composition than has been previously possible. CUVIS will provide new spectral clues to the UV absorber(s) located in the upper cloud deck that are responsible for absorbing half of the solar radiation received by Venus. With its hyperspectral imaging capability, CUVIS enables correlation between cloud features, structure, and chemistry in the upper cloud deck. CUVIS will image Venus in full sun during each of the two DAVINCI mission flybys.

The DAVINCI payload instruments will work together to comprehensively investigate the Venus environment. Table 4 summarizes how the DAVINCI instruments will address the mission's Key Questions introduced in Section 1.

Table 4. DAVINCI measurements taken with its suite of seven instruments will address key DAVINCI objectives introduced in Section 1.

| **DAVINCI Key Questions** | **DAVINCI Measurements** |
|---|---|
| What is the origin of Venus's atmosphere, and how has it evolved? Was there an early ocean on Venus, and if so, when and where did it go? How and why is Venus different than (or similar to) Earth, Mars, and exo-Venuses? | VMS determines noble gas abundance and isotope ratios to test current hypotheses of origin and evolution. CUVIS and VISOR track UV absorbers and clouds, respectively, in the upper atmosphere, and their dynamics on flybys. Both VTLS and VMS address exotic chemistry. Measurement precision of D/H by VMS and VTLS is sufficient to test the history of water. |
| Is there any current volcanism and what is rate of volcanic activity? How does the atmosphere interact with the surface? What | VMS, VTLS, and VASI work in concert to measure key trace gases near the surface and their atmospheric context, and oxygen |
| are the chemical and physical processes in the clouds and sub-cloud atmosphere? | is measured by VfOx down to the surface. Measurements and of radioactive decay products determine both the long-term average volcanism rate and the geologically recent volcanism rate. |
| Are there any signs of past processes in surface morphology and reflectance? How do tesserae compare with other major highlands and lowlands? | VenDI images reveal morphology, composition, and weathering states of representative tesserae and pave the way for future surface exploration. VenDI evaluates IR emissivity of tesserae for composition at scales of 5-200 m. Flyby VISOR 1 $\mu$m images constrain regional composition of diverse geological features at ~100 km resolution. |

## 3. Connecting Venus to Exploration beyond the Solar System

Venus is important to study not only as a deeply mysterious and compelling world of our solar system, but also as an example of a larger class of exo-Venus worlds that will likely be observed beyond the solar system in the upcoming era of the James Webb Space Telescope (JWST). Almost 5,000 exoplanets have been detected over the past several decades through a multitude of efforts. Some of these worlds will soon be observed by JWST, successfully launched in December 2021 with an anticipated mission lifetime greater than 10 years. If DAVINCI launches in 2029 and arrives at Venus in June 2031, there may be of overlap between these two missions, potentially permitting an interplay between DAVINCI in situ measurements and JWST targeted observations of exoplanets.

Exoplanets that receive Venus-like insolation levels likely represent the most observable class of terrestrial exoplanets to JWST (Kane et al. 2014). Yet these worlds will be challenging targets to interpret: most of the mass of the Venus atmosphere resides beneath its thick cloud and haze layers, but the transit transmission observations available to JWST cannot penetrate below cloud and haze and will therefore be limited to skimming the rarefied upper atmospheres of these worlds if they are enshrouded like Venus. Consequently, it has been suggested that a planet with a high altitude cloud layer could appear spectrally similar to a very different kind of planet with a thin, clear sky atmosphere (Lustig-Yaeger et al. 2019). Statistical trends in observations of such worlds could produce a “mirage” of the cosmic shoreline, the empirical dividing line in insolation-escape velocity space that separates planets with and without atmospheres (Zahnle & Catling 2017). Efficient atmospheric escape processes driven by stellar energy can erode atmospheres of planets orbiting close to their stars, producing increasingly thinner atmospheres at smaller semi-major axes. Nevertheless, the predicted decrease in cloud-top pressure at smaller semi-major axes for planets with thick, Venus-like atmospheres can produce the same apparent trend in observational data. Data from the Venus atmospheric column will help validate and constrain models that can help break this apparent degeneracy. For example, models suggest thermal phase curves could reveal the presence or absence of a thick Venus-like atmosphere, and statistical trends in populations of planets with different insolations could be compared to theoretical behavior predicted from models (Lustig-Yaeger et al. 2019). Additionally, DAVINCI’s first Venus flyby in January 2030 and resulting UV spectroscopy at 0.20 nm spectral resolution (CUVIS) may identify specific upper atmospheric chemistries for JWST to target in above-cloud transit observations of Venus-like analogues (Jessup et al. 2020).

In a more general sense, given the challenges inherent to exoplanet observations, which will typically have large error bars in even the best case scenarios for near-term observations, the worlds of the solar system including Venus provide valuable “ground truth” to improve our models and interpretations of these distant worlds. Given the particular challenges associated with observing cloudy Venus-like worlds (e.g. Barstow et al. 2016), and given that multiple potential exo-Venus planets at varied ages and stages of evolution are some of the highest priority targets for JWST (e.g., Ostberg & Kane 2019; Lustig-Yaeger et al. 2019), DAVINCI offers an opportunity for definitive “atmosphere truth” to inform and constrain studies of Venus-like exoplanets. For instance, planets of the TRAPPIST-1 system will represent a core community observation initiative with JWST (Gillon et al. 2020), and more than one of these worlds may be Venus-like (e.g., Lincowski at al. 2018; Moran et al. 2018). Furthermore, if Venus was habitable in the past, some exo-Venus planets may likewise host habitable conditions, so understanding the mechanisms and processes that governed and enabled past Venus habitability may help us to better understand the parameter space in which habitable worlds may be found beyond the solar system, allowing refinement of the habitable zone. Indeed, the inner edge of the classical habitable zone is typically used as a barometer of terrestrial planet habitability limits, as applied to other solar systems, based on our limited knowledge of Venus’s evolutionary history (e.g. Kasting et al. 1993; Kopparapu et al. 2013). Thus, improvement in our understanding of the current and past chemical and physical states of Venus represents arguably the highest priority synergistic target between the solar system and exoplanet communities for the coming years (Kane et al. 2021).

Venus may even help us to better understand how to search for and interpret oxygen as a biosignature (i.e. a remotely observable sign of life) in certain exoplanet atmospheres (e.g. Meadows 2017). Venus currently generates abiotic oxygen through $CO_2$ photolysis, which can be observed through airglow of excited ($a\ ^1\Delta_g$) oxygen on the Venusian nightside at 1.27 mm (Crisp et al. 1996), but the abundance of ground-state oxygen in the Venus atmosphere is highly unconstrained, suggesting rapid removal through chemical processes that can be better understood through DAVINCI measurements of oxygen-bearing species. Additionally, if Venus lost oceans of water to space the past, oxygen would have been generated through the processes of $H_2O$ photolysis, but this oxygen is not observed in the Venus atmosphere today. Exoplanets that lose multiple Earth oceans-worth of water could generate 100s to even 1000s of bars of abiotic $O_2$ through this process (e.g. Luger& Barnes 2015). Understanding the fate of oxygen due to possible past water loss on Venus may help to evaluate the plausibility of such models. These so-called oxygen “false positives” may be particularly relevant to JWST targets because the high activity levels and particular evolutionary histories of the low mass stars JWST will target make them especially vulnerable to generating abiotic oxygen through these processes (e.g. Meadows 2017; Meadows et al. 2018).

Beyond JWST, the Astronomy and Astrophysics 2020 decadal survey (NAS 2021) recently recommended a large infrared/optical/ultraviolet flagship observatory capable of observing exoplanets directly in reflected light around sun-like stars. Such a telescope would be capable of observing Venus-like planets in solar systems with evolutionary histories that may be similar to our own. *Pathways to Discovery in Astronomy and Astrophysics for the 2020s* discusses that observations of young Venus analog planets orbiting sun-like stars could help us understand how Venus evolved in our solar system.

## 4. Conclusions

The DAVINCI mission concept builds upon the flyby, landed, and orbital mapping missions of the past (e.g., *PVLP, Venera, Vega, Magellan, Venus Express,* and *Akatsuki*) to take the next critical step in Venus exploration: a sophisticated descent sphere-flyby combination (Figure 1). DAVINCI will deliver a chemical laboratory capable of revealing the atmospheric chemistry, a descent imager surpassing previous similar instruments on Mars (e.g., with composition and topography), an environmental package to establish context, and flyby imaging (and communications) to connect remote sensing to *in situ* exploration. The discoveries to be made by DAVINCI will close long-standing gaps in models of atmospheric evolution, Venus's water loss, and surface-atmosphere interactions. There are multiple competing models for the state of early Venus (e.g. Way et al. 2016; Turbet et al. 2021), and a precise measurement of the bulk atmosphere D/H is essential for quantifying the timing and quantity of possible water loss on Venus. Additional information will come from DAVINCI's measurements of the rock types of the tesserae and precise measurements of noble gases, which will provide multiple lines of evidence for interpreting our neighboring planet's ancient history. The resulting model inputs and constraints would benefit a broad community of next-generation scientists to understand how planetary habitability may evolve (Seager at al. 2021; Sousa-Silva et al. 2020; Greaves at al. 2020; Encrenaz et al. 2020) and to pave the way for exoplanetary modeling, observations, and exploration of Venus-like worlds beyond our solar system.

## Acknowledgments

The authors gratefully acknowledge Phase A and Phase B funding support from the NASA *Discovery* Program, as well as concept development and IRAD effort support from the NASA Goddard Space Flight Center and key partners at Lockheed Martin, Malin Space Science Systems, NASA JPL, and others. A portion of this work was carried out at the Jet Propulsion Laboratory, California Institute of Technology, under a contract with the National Aeronautics and Space Administration (80NM0018D0004). Numerous useful contributions and conversations with colleagues at Lockheed-Martin, NASA Langley Research Center, and Johns Hopkins University Applied Physics Laboratory are acknowledged by the authors. We are appreciative of the support from Lindsay Hays, Andrea Riley, Brad Zavdosky, Tiffany Morgan, and Thomas Wagner. The authors also gratefully acknowledge concept development contributions from colleagues at NASA Goddard Space Flight Center, including Martin Houghton, David Everett, Steve Tompkins, Julie Breed, Michael Amato, and Brent Robertson. Longstanding support from NASA officials including Lori Glaze, Chris Scolese, Dennis Andrucyk, Christyl Johnson, and Anne Kinney are gratefully acknowledged, as well as the inspiration of Noel Hinners and Sally Ride (deceased).

**Acronyms List**

AMPR: Above Mean Planetary Radius

DAVINCI: Deep Atmosphere Venus Investigation of Noble Gases, Chemistry, and Imaging

D/H: Deuterium to Hydrogen ratio
DS: Descent Sphere
DSN: Deep Space Network
ECAM: Engineering Camera
ES: Entry System
ESA: European Space Agency
ETU: Engineering Test Unit
GSFC: Goddard Space Flight Center
JPL: Jet Propulsion Laboratory
JWST: James Webb Space Telescope
LM: Lockheed Martin
MPR: Mean Planetary Radius
MSL: Mars Science Laboratory
MSSS: Malin Space Science Systems
NIR: Near infrared
PLVP: Pioneer Venus Large Probe
PFS: Probe Flight System
QMS: Quadrupole Mass Spectrometer
SAM: Sample Analysis at Mars
SC: Spacecraft
SfM: Structure from Motion
UV: ultraviolet
VASI: Venus Atmospheric Structure Investigation
VenDI: Venus Descent Imager
VEXAG: Venus Exploration Analysis Group
VISOR: Venus Imaging System for Observational Reconnaissance
VMS: Venus Mass Spectrometer
VTLS: Venus Tunable Laser Spectrometer